\documentclass[journal]{IEEEtran}
\usepackage{fancyhdr}
\usepackage{epsfig}
\usepackage{threeparttable}
\usepackage{epsf,epsfig}
\usepackage{amsthm}
\usepackage{amsmath}
\usepackage{amssymb}
\usepackage{amsfonts}
\usepackage[noadjust]{cite}
\usepackage{dsfont}
\usepackage{subfigure}
\usepackage{color}
\usepackage{soul}
\usepackage{algorithmic,algorithm}

\newtheorem{lemma}{Lemma}
\newtheorem{corollary}{Corollary}

\newtheorem{proposition}{Proposition}

\newtheorem{assumption}{Assumption}
\newtheorem{example}{\bf Example}

\def\qed{$\Box$}

\def\proof{\noindent{\emph{Proof:} }}

\def
\endproof{\hspace*{\fill}~\qed
\par
\endtrivlist\unskip}
\def\endproof{\hspace*{\fill}~\qed\par\endtrivlist\vskip3pt}

\def\phi{\varphi}

\def\({\left(}
\def\){\right)}

\def\bs{{\mathbf{s}}}

\def\bv{{\mathbf{v}}}

\def\by{{\mathbf{y}}}

\def\b0{{\mathbf{0}}}

\def\bS{{\mathbf{S}}}

\newcommand{\diag}{\mathrm{diag}}

\newtheorem {Remark}{Remark}

\begin{document}	
\title{\huge Analog Spatial Cancellation for Tackling the Near-Far Problem in Wirelessly Powered Communications}
\author{Guangxu Zhu, ~\IEEEmembership{Student Member,~IEEE,}, and Kaibin Huang, ~\IEEEmembership{Senior Member,~IEEE,}   \thanks{G. Zhu and K. Huang are with the Dept. of Electrical and Electronic Engineering at The  University of  Hong Kong, Hong Kong (Email: huangkb@eee.hku.hk). }}
\maketitle


\begin{abstract}
 The implementation of wireless power transfer in wireless communication systems opens up a new research area, known as wirelessly powered communications (WPC). In next-generation heterogeneous networks where ultra-dense small-cell base stations are deployed, simultaneous-wireless-information-and-power-transfer (SWIPT) is feasible over short ranges. One challenge for designing a WPC system is the severe near-far problem where a user attempts to decode an information-transfer (IT) signal in the presence of extremely strong SWIPT signals. Jointly quantizing the mixed signals causes the IT signal to be completely corrupted by quantization noise and thus the SWIPT signals have to be suppressed in the analog domain. This motivates the design of a framework in this paper for analog spatial cancellation in a multi-antenna WPC system. In the framework, an analog circuit consisting of simple phase shifters and adders, is adapted to cancel the SWIPT signals by multiplying it with a cancellation matrix having unit-modulus elements and full rank, where the full rank retains the spatial-multiplexing gain of the IT channel. The unit-modulus constraints render the conventional zero-forcing method unsuitable. Therefore, the paper presents a novel systematic approach for constructing cancellation matrices. For the single-SWIPT-interferer case, the matrices are obtained as truncated Fourier/Hadamard matrices after compensating for propagation phase shifts over the SWIPT channel. For the more challenging multiple-SWIPT-interferer case, it is proposed that each row of the cancellation matrix is constructed as a Kronecker-product of component vectors, with each component vectors designed to null the signal from a corresponding SWIPT interferer similarly as in the preceding case.
\end{abstract}



\section{Introduction}
Recent years have seen a series of breakthroughs  in wireless communication technologies, such as millimeter wave communications, massive \emph{multiple-input-and-multiple-output} (MIMO), and small-cell networks, which jointly provide a solution for coping with exponential growth of mobile data traffic. In contrast, wireless power transfer (WPT) using microwaves remains a relatively stagnant field and the current low transfer efficiencies due to severe propagation  loss prevents its extensive  commercialization. This key challenge, however, may be tackled by implementing WPT using next-generation wireless networks as a platform where the increasing network densification reduces the transmission distances to merely tens of meters \cite{Lopez:utradensesmallcell} and the deployment of large-scale arrays enable sharp beamforming to suppress dispersion of radiated energy \cite{xiaomingchen:SWIPTmultiantenna}. This vision has motivated active research on seamless integration between WPT and wireless communications, opening a new area called \emph{wirelessly powered communications} (WPC). In WPC networks, the ranges of power transfer (PT) (e.g., tens of meters) and information transfer (IT) (up to several kilometers) can be drastically different, leading to a severe \emph{near-far} problem \cite{Huang:CuttingLastWiress:2014}. This results in the coexisting of PT and IT signals with the power difference of many orders of magnitude. 
{
Note that a practical issue incurred by the near-far problem is that the joint quantization of the strong and weak signals at the latter to be corrupted by quantization noise. This issue is particularly severe in the context of WPC system compared with that in the conventional  uplink multi-user access scenario, since the power difference between the PT and IT signals could be many orders of higher than that between the IT signals from different users \cite{Huang:CuttingLastWiress:2014}. The solution for this issue, to the best of the authors' knowledge, has not been reported in the existing WPC literature.}
This paper addresses this issue by  presenting a framework for spatial cancellation of PT signals using an analog circuit prior to quantization, referred to as \emph{analog spatial cancellation}.

\subsection{Prior Work}

For short-range transmission, the same carrier can be used for both PT an IT, which is commonly known as \emph{simultaneous wireless information and power transfer} (SWIPT). The idea was first explored in \cite{Varshney:TransportInformationEnergy:2008,GroverSahai:ShannonTeslaWlssInfoPowerTransfer} from the information theoretic perspective and  the fundamental tradeoff between harvested energy and information capacity in a SWIPT system is quantified. The results are based on an ideal assumption that the receiver is able to harvest energy and decode information from the same received signal. The difficulty of realizing  this assumption in practice motivated the design of power-splitting SWITP receiver  in \cite{Zhang:MIMOBCWirelessInfoPowerTransfer} where the received signal is split for separate energy harvesting and information decoding  and the rate-energy tradeoff of a multiuser MIMO SWIPT system was characterized based on this receiver architecture. The idea of SWIPT opens up a rich set of interesting research opportunities  having a similar theme of revamping communication theory and techniques to incorporate the feature of WPT.  Various types of SWIPT systems have been proposed and studied recently including broadband SWIPT      \cite{HuangLarsson:SIPTBroadbandChannel,NgLo:MultiuserOFDMSInfoPowerTransfer}, relay-assisted SWIPT \cite{nasir:RelayHarveing:2013,zhiguoding:powerallocation,caijunzhong:SWIPTfullduplex}, cognitive SWIPT  networks \cite{Niyato:cognitiveWPC,DWKNg:multiobjectiveresourceallocation}, and interference channels with SWIPT \cite{Guangxu:SWIPTrelay,Bruno:SWIPTinterference}.   {In future dense heterogeneous networks, SWIPT  links will coexist with much weaker IT links, creating the mentioned near-far problem \cite{Huang:CuttingLastWiress:2014}.}  Surveys of latest  advancements in this active area can be found in \cite{ulukus2015energy,krikidis2014simultaneous,S.Bi:WPCmag,NiyatoandD.I.Kim:WPTsurvey}.

Besides SWIPT, other configurations of WPC systems are also developed in the literature. A  WPC network  was proposed in \cite{Zhang:ThputMaxWirelessPowerCommNet} where base stations power uplink  mobiles by  downlink PT. The throughput maximization problem was formulated and solved in this paper. The work has  been extended to multi-antenna system with energy beamforming \cite{ruizhang:multiantennaWPC}, full-duplex systems \cite{C.K.Ho:fullduplexWPC}, massive MIMO systems  \cite{G.Yang:massiveMIMOWPC} and large-scale communication networks \cite{ruizhang:WPCnetworks}. However, it is impractical to rely on only existing base stations for achieving network coverage of PT as current inter-cell distances are much longer than PT ranges. One practical solution as proposed in  \cite{HuangLauArXiv:EnablingWPTinCellularNetworks:2013} is to densely deploy power stations dedicated for PT, called  \emph{power beacons}. power beacons have low complexity and require no backhaul, allowing dense deployment to increase PT coverage. Moreover, those power beacons with Internet access can double as ultra-dense small-cell BSs. 

{ 
In addition to the aforementioned design of WPC systems and techniques, recent information theoretic research building on the initial work in \cite{Varshney:TransportInformationEnergy:2008,GroverSahai:ShannonTeslaWlssInfoPowerTransfer} has been reported in \cite{shaviv2016capacity,tandon2014code,fouladgar2014constrained}. In \cite{shaviv2016capacity}, it was shown that by exploiting the channel state information available at the wireless charger, the capacity of the WPC system can be significantly improved by performing transmit-power adaption, and the fundamental capacity limit under different levels of side information sharing was characterized from the information theoretic perspective. On the other hand, coding schemes have been designed in \cite{tandon2014code,fouladgar2014constrained} to optimize the tradeoffs between the IT rate and the PT efficiency. Despite this theoretical research, transforming WPC from theory to practice still faces many unsolved practical issues, and the said near-far problem remains one of the key challenges for designing WPC systems.
}

As mentioned, given the scale of their  power difference, jointly quantizing  the received  PT and IT signals renders the latter completely corrupted by quantization noise and it is impractical to solve this problem by increasing the ADC resolution. The conventional approaches to avoid this problem is to  perform PT and IT in separate frequency sub-channels (see e.g., \cite{HuangLarsson:SIPTBroadbandChannel}) or by time sharing  \cite{nasir2015TCOM} but they are not without drawbacks.  Time shared  PT-and-IT reduces their efficiency/rate and furthermore requires strict synchronization between users. For frequency division PT-and-IT, the suppression of ultra-strong PT signal at an information decoder requires a sharp analog band-pass filter plus sufficiently large frequency separation between PT and IT signals. SWIPT using the same spectrum does not have the drawbacks mentioned above but requires analog spatial cancellation  of the IT signal prior to quantizing the  PT signal, which is a largely uncharted area and the theme of the paper.

\subsection{Contributions and Organization}

We  consider WPC system where a macrocell BS, called a \emph{IT BS},  performing IT to a single user who also receives strong intended/unintended SWIPT signals transmitted by ultra-dense small-cell BSs (or power beacons with Internet access), called \emph{SWIPT BSs}.  All nodes are equipped with multi-antennas. On one hand, the short-range SWIPT with sharp beamforming over sparse scattering is  modeled as a free-space channel. The SWIPT BSs are assumed to have different angles-of-arrival at the user. { It is worth pointing out free-space channels are essential for efficient WPT and thus widely assumed in the literature of WPT (see e.g., \cite{Brown:RadioWPTHistory:1984, N.Shinohara:WPTviaradiowaves,Huang:CuttingLastWiress:2014}).}  On the other hand, the long-range IT channel with rich scattering is modeled as an independent and identically distributed (i.i.d.) Rayleigh fading channel. This enables spatial multiplexing over the MIMO IT channel.  Our work focuses on the user's decoding  of  the data streams in the IT signal. In addition, besides harvesting energy from the SWIPT signal, the user can also decode information in the signal if it is intended for the user. The near-far problem is irrelevant for the processing of the SWIPT signal that is thus neglected in our work.

In this paper, we identify the said near-far problem and present a novel framework of analog spatial cancellation implemented at the user for suppressing the strong SWIPT signal prior to quantizing the IT signal. The operation is implemented  using an analog circuit  comprising simple RF components including phase shifters and adders, which is attached  to the receive antenna array.  The circuit implements multiplication of the observation vector of receive antennas by a matrix with unit-modulus elements, which is called a \emph{cancellation matrix} and whose rows \emph{cancellation vectors}.  The design problem is  formulated as the optimization of maximizing the row rank of the cancellation matrix  under zero-forcing (ZF) constraints for nulling the SWIPT signals and  unit-modulus constraints for individual elements, where rank maximization maximizes  the spatial multiplexing gain for IT.  The unit-modulus constraints render the conventional ZF cancellation technique based on linear algebra inapplicable. The focus of the work is to develop a systematic approach for solving the design problem. Essentially, the approach finds a  set of independent cancellation  vectors, each of which is orthogonal to and thus cancel the set of SWIPT channel vectors. The independence between the vectors ensures that the cancellation matrix has full row rank.  The contributions are summarized as follows. 

\begin{itemize}

\item Consider the simple case of single SWIPT BS. The proposed design of the analog cancellation circuit, which solves the design problem,  comprises two sequential parts. The first part performs phase de-rotation to compensate for propagation phase shifts  of observations from different  receive antennas, reducing the SWIPT channel to be equivalent to an all-one vector. The second part transforms the mixed input signal by a truncated Fourier/Hadamard matrix with the all-one row eliminated.  Leveraging the row orthogonality of the matrix, the signal from the effective SWIPT channel (an all-one vector) is nulled. The unit-modulus properties of the Fourier/Hadamard matrix elements facilitate analog implementation of the transform using phase shifters. Furthermore, the full row rank of the transform matrix ensures maximum spatial multiplexing gain of the IT signal under the SWIPT interference cancellation constraints.

\item Consider the general and more complex case of  multiple SWIPT BSs. The preceding design cannot be easily extended to cancel multiple SWIPT signals arriving from different angles. A more sophisticated design   is proposed for the current case. Specifically,  by exploiting its  Vandermonde structure,  each free-space SWIPT channel can be  decomposed  into the Kronecker product of component phase-shift vectors. This motivates the proposed construction of each individual cancellation vector as a Kronecker product of component phase-shift vectors. The orthogonality between an arbitrary pair of cancellation and SWIPT channel vectors can be achieved by   the orthogonality between  any pair of their components vectors. According to this property,  a single cancellation vector can be constructed, using a Fourier-based construction method, to null all multiple SWIPT interference channels based on a mapping between the components of the former to targeted SWIPT channels.
Repeating the construction for all possible mappings generates a mother set of cancellation vectors. Then a subset of independent vector can be selected from the mother set by a greedy/random search, yielding the desired cancellation matrix solving the design problem for the current case.

\end{itemize}

The remainder of the paper is organized as follows: Section II introduces the system model. 
Section III formulates the design problem for implementing the analog spatial cancellation. Two simple but optimal solutions based on Fourier or Hadamard matrices targeting for the single SWIPT BS case are proposed in Section IV. Then, a novel systematic construction framework tackling the more challenging multiple SWIPT BSs case is developed in Section V and VI for two different sub-cases. Numerical results and discussions are presented in Section VII, followed by conclusion in Section VIII. 

\section{System Model}
\begin{figure}[t!]
\centering
\includegraphics[width = 8.5cm]{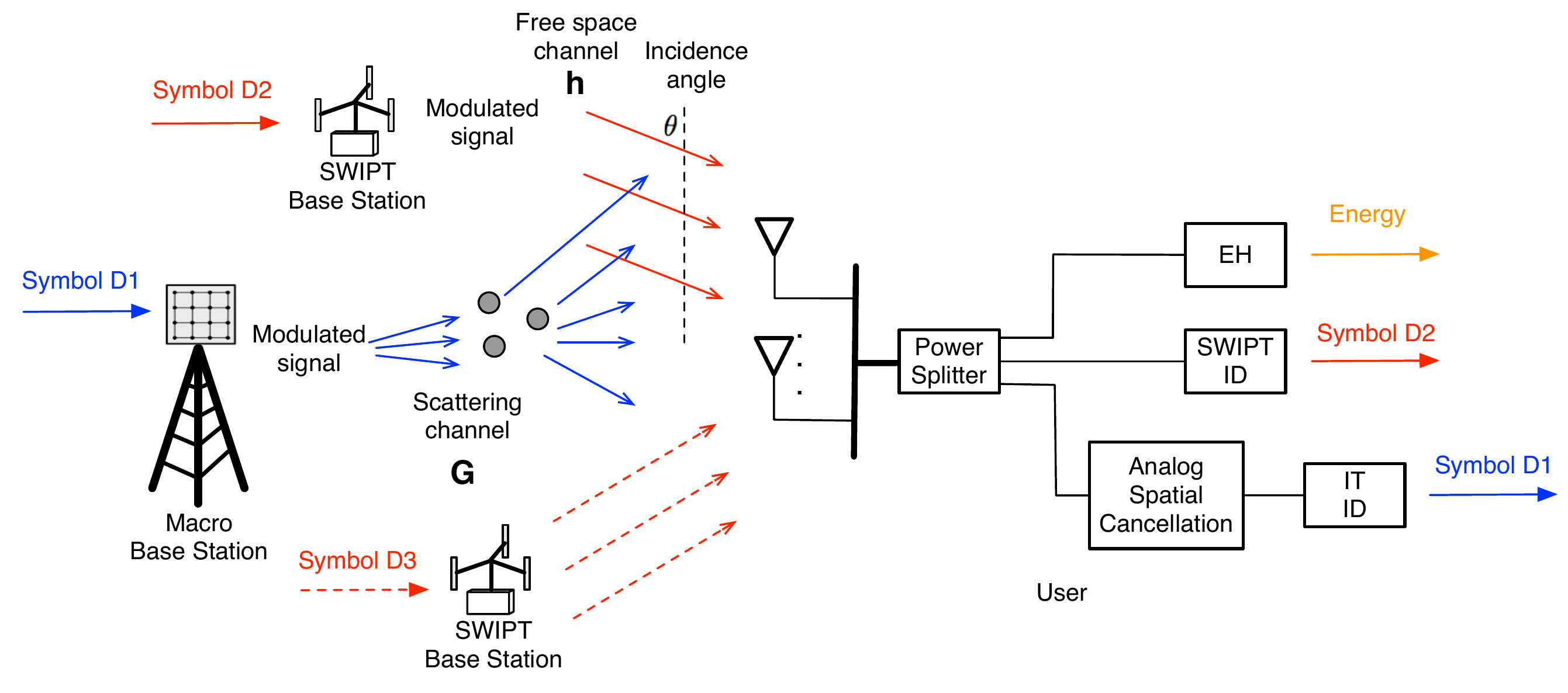}
\caption{A WPC system where IT is interfered with by an intended and an unintended SWIPT BS. The blue solid arrow represents the IT signal, the red solid arrow denotes the intended SWIPT signal (which will be decoded by the user) and the red dash arrow stands for the unintended SWIPT interferer (which will not be decoded by the user).}\label{fig:1}
\end{figure}

As shown  in Fig. \ref{fig:1}, we consider a WPC system where a user attempts to retrieve spatial multiplexed data streams transmitted by an IT BS (marco BS) over a MIMO channel, in the presence of strong  SWIPT signals beamed by  $K$ SWIPT BSs ($K=2$ in Fig. \ref{fig:1}). The SWIPT signals can be intended or unintended to the considered user. It is assumed that the IT BS is equipped with $N_t$ antennas and the user is equipped with a linear array of $N_r$ antennas, while all the SWIPT BSs are provisioned  with arrays for free-space beamforming. 
 \begin{assumption}\emph{The short-range SWIPTs  are over narrow-band \emph{free-space} channels\footnote{Free-space beamforming for SWIPT, essential for high WPT efficiency, is possible due to sparse scattering in the short-range channel between the PB and mobile.}, while the long-range narrow-band IT channel in the same bandwidth is characterized by \emph{rich scattering} modeled as i.i.d. Rayleigh fading.} 
 \end{assumption}

\noindent Based on this assumption, a SWIPT channel can support only a single data stream, while up to $\min(N_t, N_r)$ streams can be spatially multiplexed in the IT channel. Let $N_r \times 1$ vector ${\bf h}_i$ represents the effective SIMO channel (after transmit beamforming) between the $i$th SWIPT BS and the user.  The signal received at the user can be represented by a $N_r\times 1$ vector $\by$ given as 
\begin{align}
{\bf y} = {\bf G}{\bf x} + \sum_{i=1}^{K}{\bf h}_is_i + {\bf n},
\end{align}
where the $N_r\times N_t$ matrix ${\bf G}$ represents the IT channel,  ${\bf x} \in \mathbb{C}^{N_t \times 1}$ denotes the IT signal transmitted by the IT BS,  $s_i$ is the signal transmitted by the $i$-th SWIPT BS, and ${\bf n}\in \mathbb{C}^{N_r \times 1}$ represents the additive white Gaussian noise. 
Assuming that the incident SWIPT signal from the $i$-th SWIPT BS can be approximated as a plane wave with the angle-of-arrival $\theta_i$, the effective channel vector ${\bf h}_i$ can be written as ${\bf h}_i=a_i {\bf v}(\Theta_i)$ where scalar $a_i$ captures the path loss as well as the beamforming gain of ${\bf h}_i$, and $N_r \times 1$ vector ${\bf v}(\Theta_i)$ represents the phase response for the linear receive array. Specifically,\begin{align}\label{sys:1}
{{\bf v}(\Theta_i)} =[1,e^{j\Theta_i},\cdots,e^{j(N_r-1)\Theta_i}]^T,
\end{align}
where   $\Theta_i = \frac{2\pi d}{\lambda}\cos\theta_i$ denotes the constant phase difference  between the signal observed by two adjacent receive antennas with $d$ being the antenna separation distance and $\lambda$ representing the carrier wavelength. The received signal $\bf y$ is split for energy harvesting (EH), SWIPT information decoding (ID) and IT ID as illustrated in Fig.~\ref{fig:1}.



\section{Problem Formulation}\label{sec:PF}
\begin{figure}[t!]
\centering
\includegraphics[width = 8cm]{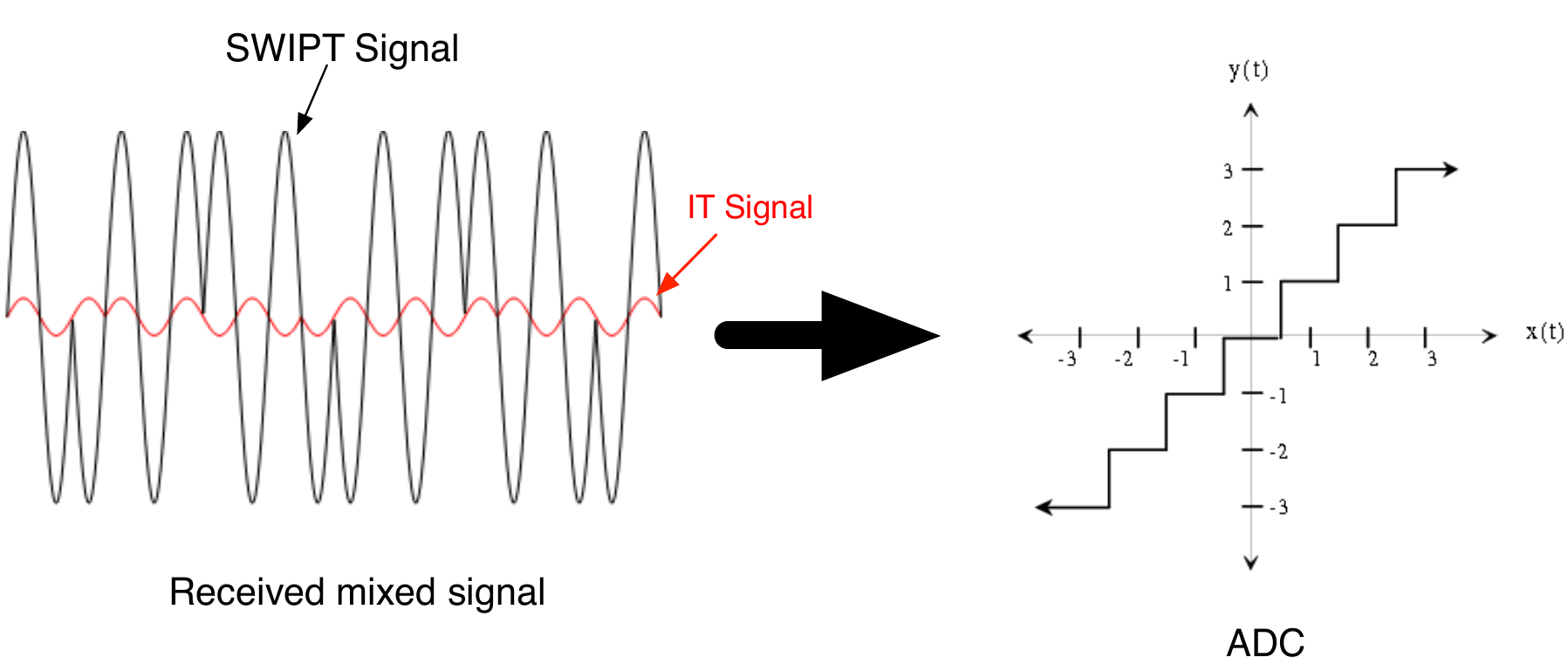}
\caption{Quantization of the received signal without analog spatial cancellation.}\label{fig:ADC}
\end{figure}

The mentioned severe near-far problem leads to a extremely low SQNR when quantizing the mixed SWIPT-IT signal, making it difficult if not impossible to decode the weak IT signal. Specifically, in the quantization process illustrated in Fig. \ref{fig:ADC}, the SWIPT signal is scaled to span the full dynamic range of the ADC. This
reduces the peak magnitude of IT signal to be smaller than the quantization step size. The corresponding SQNR for IT signal can be calculated as follows \cite{Proakis:DSP},
\begin{equation}\label{quan_error}
{\sf SQNR} \approx 6.02\times b + c - R,
\end{equation}
where $b$ represents the given ADC resolution in bit, $R$ denotes the power ratio between the received SWIPT and IT signals, and the constant $c = -8.5 \sim 1.76 \;{\sf dB}$ depends on the distribution of the input signal. For example, given $R = 90\;{\sf dB}$ and $b = 10$ bits, the SQNR can be computed to be approximately $-30\;{\sf dB}$ which makes it impossible to recover the data in the IT signal. Therefore, to tackle the near-far problem, it is essential to decouple the IT and SWIPT signals in the analog domain and then quantize them separately.

To reliably decode the IT signal, we design an analog circuit that implements the multiplication the received signal with the cancellation matrix denoted as ${\bf S}$ to null  the strong SWIPT signal. Under the ZF constraints, the row rank of ${\bf S}$ is maximized such that maximum multiplex gain can be achieved for the effective MIMO fading channel, defined as $\tilde{\bf G} = {\bf SG}$. The proposed simple design comprises an array of interconnected adjustable phase shifters. For the conventional digital spatial cancellation using a DSP processor, both the magnitudes and phases of signals can be varied. In contrast, for the proposed design, we can only adjust the phases of signals via phase shifters, which introduces the unit-modulus constraints to the elements of the ZF combining matrix  $\bf S$. Based on the above discussion,  the design problem is formulated as follows: 
\begin{equation}
\begin{aligned}
\mathop {\max }\limits_{\bf{S}} \; &{\sf Rank} ({\bf S})\\
{\textmd{s.t.}}\;\;&{\bf S}[{\bf h}_1,\cdots,{\bf h}_{K}] = {\bf 0}, \\
&|[{\bf S}]_{m,n}| = 1,\;\forall m,n, 
\end{aligned}
\label{P:1}
\end{equation}
where ${{\bf S} \in \mathbb{C}^{N_{\sf rank}\times N_r}}$ is the full row rank cancellation matrix needs to be designed, $N_{\sf rank}$ represents the row rank of ${\bf S}$, $[{\bf S}]_{m,n}$ is the $(m,n)th$ element of ${\bf S}$. To ensure the existence of the cancellation matrix, it is assumed that $N_r > K$.
Using  (\ref{sys:1}), the  optimization problem can be  equivalently written as 
\begin{equation}({\bf P1})\qquad 
\begin{aligned}
\mathop {\max }\limits_{\bf{S}} \; &{\sf Rank} ({\bf S})\\
{\textmd{s.t.}}\;\;&{\bf S}[{\bf v}(\Theta_1),\cdots,{\bf v}({\Theta_{K}})] = {\bf 0}, \\
&|[{\bf S}]_{m,n}| = 1,\;\forall m,n,
\end{aligned}
\label{P:2}
\end{equation}
where ${{\bf v}(\Theta_i)}$ is given in \eqref{sys:1}.

The main challenge for solving the above optimization problem is to satisfy the unit-modulus constraints, i.e., $|[{\bf S}]_{m,n}| = 1,\;\forall m,n$. In particular, the unit-modulus constraints make the feasible set, denoted as $\mathbb{S}$, for  problem P1 no longer a traditional Euclidean vector/space. The reason is that the vectors in $\mathbb{S}$ do not satisfy the \emph{closure properties} for addition and scalar multiplication, i.e., $\forall {\bf s}_1,{\bf s}_2 \in \mathbb{S}, {\bf s}_1 + {\bf s}_2 \notin \mathbb{S}$, and $\forall {\bf s}_1 \in \mathbb{S}, a \in \mathbb{C}, a{\bf s}_1 \notin \mathbb{S}$ if $|a| \neq 1$. As a result, the conventional approach of computing ZF vectors as those lying in the null space of SWIPT channel vectors, e.g., singular value decomposition (SVD), do not solve problem P1 since the elements of such vectors have different norms. Therefore, a new systematic solution  approach based on new mathematical tools is developed in the sequel.

\section{Analog Spatial Cancellation with a Single SWIPT BS}\label{sec:5}
In this section, we  consider the simple case where only a single SWIPT BS is employed to wirelessly power the user. Then, problem {P1} in (\ref{P:2}) reduces to,
\begin{equation}
\begin{aligned}\mathop {\max }\limits_{\bf{S}} \; &{\sf Rank} ({\bf S})\\
{\textmd{s.t.}}\;\;&{\bf S}{\bf v}(\Theta)= {\bf 0}, \\
&|[{\bf S}]_{m,n}| = 1,\;\forall m,n, 
\end{aligned}
\label{P:3}
\end{equation}

Note that, without the unit-modulus constraints, the maximum rank of the desired matrix $\bf S$ should be equal to the dimension of the null space of ${\bf v}(\Theta)$, i.e., $(N_r - 1)$, since one degree-of-freedom (DoF) is used to suppress the SWIPT signal. Therefore, given the constraints, it is interesting to investigate whether a rank-($N_r - 1$) solution $\bf S$ can still be obtained. To this end, 
two simple but optimal analog spatial cancellation schemes are proposed in the following sub-sections.

\subsection{The Fourier Based Scheme}\label{sec:3.1}

\subsubsection{Design}\label{sec:3.1.1}

\begin{figure}[t!]
\centering
\includegraphics[width= 8cm]{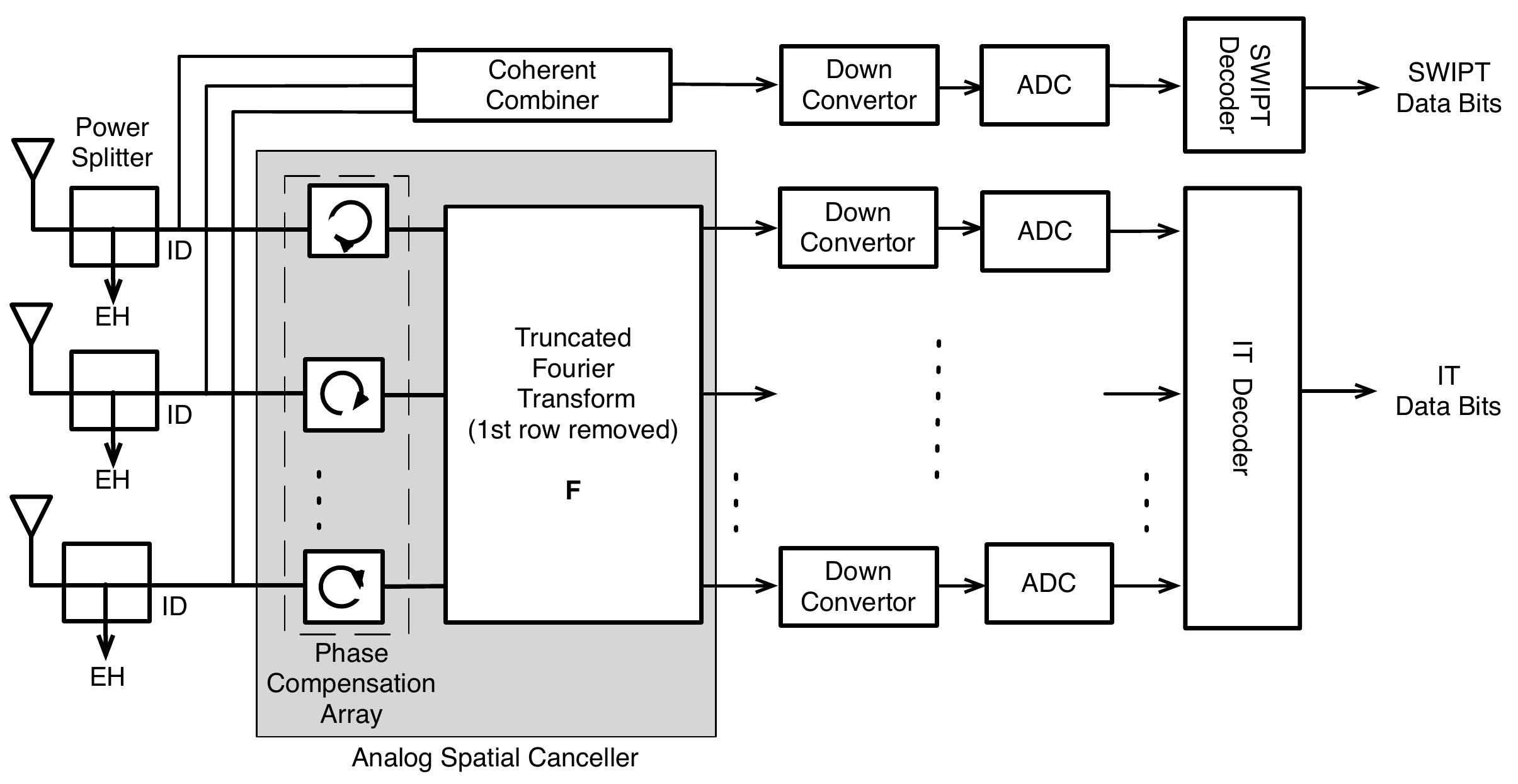}
\caption{Architecture for Fourier based analog spatial cancellation.}
\label{Fig:Fourier_scheme}
\end{figure}

As illustrated in Fig. \ref{Fig:Fourier_scheme}, for the processing of the SWIPT signal, if it is intended for the user, it can be easily decoded by using a simple coherent combiner{\footnote{Since the SWIPT signal propagate as a plane wave in the assumed free-space channel, the coherent combine of the SWIPT signal can be easily implemented in the analog domain using a phase compensation array plus an adder.}}, as the SWIPT signal is unaffected by the said near-far problem. On the other hand, for the IT signal decoding, the SWIPT interference  needs to be suppressed first at the analog domain by analog spatial cancellation. Particularly, the analog spatial canceller consists of two components, i.e., the \emph{phase compensation array} and the \emph{truncated Fourier transform}. Given the knowledge of the angle-of-arrival $\theta$, the phase compensation array aligns the phases of the received SWIPT signals at different antennas such that the received SWIPT symbol is multiplied by a scaled all-one vector. To be specific, the phase compensation array can be expressed as ${\bf R} = \diag (1,e^{-j\Theta},\cdots,e^{-j(N_r-1)\Theta})$, thus, the phase-compensated signal, denoted by ${\by}_c$, can be given by
\begin{align}
{{\bf y}_c} &= {\bf Ry} = {\bf RGx}+{\bf Rh}_1s_1 + {\bf Rn} \notag\\
&= {\bf RGx} + a_1{\bf u}s_1 + {\bf Rn},
\end{align}
where ${\bf u} = [1,1,\cdots,1]^T$. 

Next, the truncated Fourier transform multiplies the input with a truncated Fourier matrix with the first row removed:
 \begin{align}
{\bf F} = \left[ {\begin{array}{*{20}{c}}
1&w&{{w^2}}& \cdots &{{w^{{N_r} - 1}}}\\
1&{{w^2}}&{{w^4}}& \cdots &{{w^{2({N_r} - 1)}}}\\
 \vdots & \vdots & \vdots &{}& \vdots \\
1&{{w^{{N_r} - 1}}}&{{w^{2({N_r} - 1)}}}& \cdots &{{w^{({N_r} - 1)({N_r} - 1)}}}
\end{array}} \right], \notag
\end{align}
where $w = e^{-j2\pi/N_r}$. 
Since the rows of the Fourier matrix are orthogonal, the multiplication suppresses the strong SWIPT signal and the IT signal will be extracted from the mixed signal for further decoding.

Mathematically, the transformed signals, denoted as ${\bf y}_t$, is 
\begin{align}\label{FT:1}
{{\bf y}_t} &= {\bf Fy}_c = {\bf FRGx}+ a_1{\bf Fu}s_1 + {\bf FRn}\notag\\
&= {\bf FRGx} + {\bf FRn}.
\end{align}
{Note that the all-one column vector $\bf u$ is exactly the transpose of the first row in the Fourier matrix, yielding  ${\bf Fu = 0}$ in (\ref{FT:1}) due to the orthogonality between rows of a Fourier matrix.}  This suppresses the SWIPT signal at the output of the truncated Fourier transform.

 After phase compensation and truncated Fourier transform, the equivalent MIMO fading channel for the IT signal is given by $\tilde {\bf G} = {\bf FRG}$, which is a rank-$(N_r-1)$ matrix. It means that, with one DoF used to suppress the SWIPT signal, the remaining  ${(N_r - 1)}$ DoF can still be exploited in the effective MIMO fading channel for spatial multiplexing. Note that SWIPT channel can provide one DoF for data transmission, which interestingly implies that maximum multiplexing gain of ${N_r}$ can be achieved in the considered system with analog spatial cancellation.
 
The above discussion leads to the following main result.

\begin{proposition}\label{prop:0}
The solution of the optimization problem in (\ref{P:3}) can be obtained as follows:

\begin{align}\label{solution:1}
&{\bf S}_F  = {\bf FR} =\notag\\
&\left[ {\begin{array}{cccc}
1&w& \cdots &{{w^{{N_r} - 1}}}\\
z&z{{w^2}}& \cdots &z{{w^{2({N_r} - 1)}}}\\
 \vdots & \vdots &{}& \vdots \\
z^{N_r - 1}&z^{N_r - 1}{{w^{{N_r} - 1}}}&\cdots &z^{N_r - 1}{{w^{({N_r} - 1)^2}}}
\end{array}} \right]\!\!,
\end{align}
where $z = e^{-j\Theta}$.

{
\proof
According to the preceding analysis presented above, the optimal analog cancellation matrix can be obtained as a truncated Fourier matrix ${\bf F}$ times a phase compensation matrix ${\bf R}$ as shown in (\ref{solution:1}). The optimality of  (\ref{solution:1}) can be proven as follows. Firstly, note that the phase compensation matrix is an unitary matrix and the truncated Fourier matrix has a rank of $(N_r - 1)$. Thus, it is easy to verified that ${\sf rank}({{\bf S}_F}) = N_r - 1$ which achieves the desired maximum rank. Next, as shown in (\ref{FT:1}), we have ${{\bf S}_F}{\bf v}(\Theta) = {\bf FRv}(\Theta) = {a_1\bf Fu} = {\bf 0}$ which enforces the zero forcing constraint in (\ref{P:3}). Finally, it can be observed from (\ref{solution:1}) that each element in ${{\bf S}_F}$ involves only phase shift and the unit-modulus constraints are also satisfied. This completes the proof.
\endproof
}
\end{proposition}

\begin{Remark}\label{remark:4}
 \emph{A close observation reveals that the proposed ${\bf S}_F \in \mathbb{C}^{(N_r-1)\times N_r}$ is a scaled para-unitary matrix, i.e., ${{\bf S}_F{\bf S}_F^{H}} = N_r{\bf I}_{N_r -1}$. In other words, the rows in ${\bf S}_F$ are mutually orthogonal and have the same norm. Such a property is quite important in the perspective of system stability, which can be indicated by the condition number. Generally, the smaller the condition number is, the more stable the system can be. Condition number of ${\bf S}_F$ can be calculated as $\textmd{cond}({\bf S}_F) = \frac{\sigma_{\sf max}({\bf S}_F)}{\sigma_{\sf min}({\bf S}_F)}$, where ${\sigma_{\sf max}({\bf S}_F)}$ and ${\sigma_{\sf max}({\bf S}_F)}$ are maximal and minimal singular values of matrix ${{\bf S}_F}$. It is easy to show that the minimal condition number of ${\bf S}_F$ is one, given the ${\bf S}_F$ is a scaled para-unitary matrix.}
\end{Remark}


%
\subsubsection{Implementation and complexity}
Since a Fourier matrix for an arbitrary $N_r$ exists, the design in Fig. \ref{Fig:Fourier_scheme} is valid for any $N_r \geq 2$. The implementation requires $(N_r - 1)$ phase shifters for the phase compensation array and $(N_r - 1)^2$
shifters for the truncated Fourier transform. As a result, the total number of the required phase shifters is  $N_r(N_r - 1)$, which may lead to a high implementation cost if $N_r$ is large. This motivates  an alternative low-cost scheme proposed in the following subsection.


\subsection{The Hadamard Based Scheme}
\subsubsection{Design}
The architecture of the Hadamard based scheme is similar to that of the Fourier based one shown in Fig. \ref{Fig:Fourier_scheme}. The only difference is that the matrix ${\bf F}$ is replaced by a  truncated Hadamard transform $\bar {{\bf H}}$, i.e., a Hadamard matrix with the first row removed.  
A Hadamard matrix is a square matrix whose entries are either $+1$ or $-1$ and whose rows are mutually orthogonal.
For example, a $4\times 4$ Hadamard matrix is 
\begin{equation}
{\bf{H}}_4  = \left[ {\begin{array}{*{20}c}
   1 & 1 & 1 & 1 \\
   1 & -1 & 1 & -1 \\
   1 & 1 & -1 & -1 \\
   1 & -1 & -1 & 1
\end{array}} \right].
\end{equation}
A necessary condition for a $n\times n$ Hadamard matrix ${\bf H}_n$ to exist is that $n$ is equal to $2$ or a positive multiple of $4$, and the corresponding constructing methods can be found in \cite{Yarlagadda:Hadamardmatrix:1997}. 

Extending Proposition \ref{prop:0} gives the following result.
\begin{proposition}
An alternative solution of the problem in (\ref{P:3}) is given by
\begin{align}\label{eq:p2}
{\bf S}_H = \bar {{\bf H}}{\bf R}.
\end{align}

{
\proof
The optimality of (\ref{eq:p2}) can be shown following the arguments in the proof of Proposition \ref{prop:0}.
\endproof
}
\end{proposition}
Remark \ref{remark:4} also applies here with ${\bf S}_F$ replaced with ${\bf S}_H$.

\subsubsection{Implementation and complexity}
\begin{figure}[t!]
\centering
\includegraphics[width = 8cm]{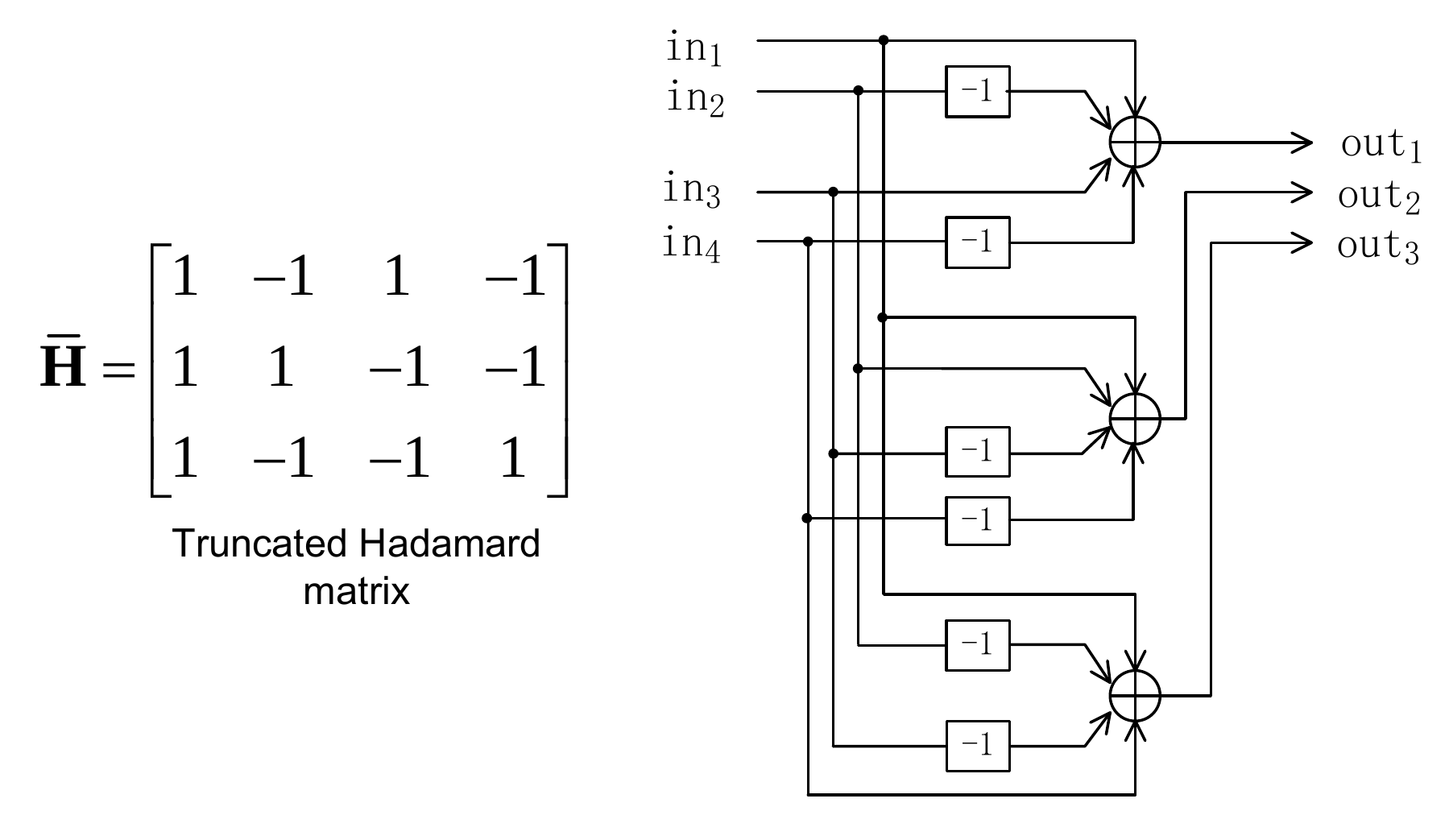}
\caption{Example circuit implementing the truncated Hadamard transform.}
\label{Fig:Hadamard_transformer}
\end{figure}

Compared with the Fourier based scheme, the drawback of the Hadamard based scheme is that $N_r$ must be $2$ or a multiple of $4$. However, the latter incurs a much lower implementation cost. In particular, since the entries of a Hadamard matrix are either $1$ or $-1$, the implementation of the  truncated Hadamard transform requires no phase shifter but just $(N_r - 1)$ adders. Fig. \ref{Fig:Hadamard_transformer} gives an example for the practical circuit of  truncated Hadamard transform with $N_r = 4$. As a result, the total number of required phase shifters reduces from $N_r(N_r-1)$ for the Fourier based design to $(N_r-1)$, corresponding to nearly $N_r$ times cost reduction. A comparison including implementation requirement and complexity for different proposed schemes are given in Table \ref{summary:table}.

\begin{table}[ht]
\centering
\caption{Comparison of the Fourier based and Hadamard based schemes}
\begin{tabular}{|p{2.25cm}|c|c|}
\hline
{} & Fourier based scheme & Hadamard based scheme  \\
\hline
Number of required receive antennas & $\forall N_r \geq 2$ & $N_r = 2$ or $4n$ \\
\hline
Number of required phase shifters  & $N_r(N_r - 1)$ & $N_r - 1$ \\
\hline
Number of required adders  & $N_r - 1$ & $N_r - 1$ \\
\hline
Rank of ${\bf S}_{F}$ or ${\bf S}_{H}$  & $N_r - 1$ & $N_r - 1$ \\
\hline
Condition number of ${\bf S}_{F}$ or ${\bf S}_{H}$ & $1$ & $1$ \\
\hline
\end{tabular}
\label{summary:table}
\end{table}

\section{Analog Spatial Cancellation with  the Maximum Number of SWIPT BSs}\label{sec:6}
In this section, we consider the case of multiple SWIPT BSs. For ease of exposition, 
let $K =  K_{\max}$  where $K_{\max}$ denotes the maximum number of SWIPT BSs such that their signals can be cancelled by the user in the analog domain and is derived in the sequel.   The design of  analog spatial cancellation presented in the preceding section for the case of single SWIPT BS cannot be directly applied to the current case since the multiple SWIPT signals with different angles-of-arrival cannot be simply aligned and cancelled using the scheme of phase compensation plus truncated Fourier/Hadamard transform. To address this issue, a more sophisticated  systematic approach, called \emph{Kronecker based construction}, for computing the phases of the phase shifters in the analog cancellation circuit (or equivalently the cancellation matrix) is proposed in the current section.


It can be observed from Problem P1 that the maximum row rank of the  cancellation matrix $\bS$ is $N_r - K$. Under the full-rank constraint, $\bS$ can be written as  ${\bf S} = [{\bf s}_1, {\bf s}_2, \cdots, {\bf s}_{N_r - K}]^T$, where $\{{\bf s}_i\}$  are  linearly independent cancellation vectors. Then problem P1  can be rewritten to explicitly reflect the design goal of constructing  $(N_r - K)$ linearly independent cancellation vectors as follows: 

\begin{equation}({\bf P2})\qquad 
\begin{aligned}
\mathop {\max }\limits_{{\bf s}_1,{\bf s}_2,\cdots,{\bf s}_{N_r - K}} \; &{\sf Rank} ({\bf S})\\
{\textmd{s.t.}}\;\;\;\;&{\bf s}_i^T[{\bf v}(\Theta_1),\cdots,{\bf v}({\Theta_{K}})] = {\bf 0},\; \forall i\\
&|[{\bf s}_i]_j| = 1, \; \forall i,j, 
\end{aligned}
\label{P:4}
\end{equation}

It is important to note that unlike traditional interference cancellation, the condition $K <  N_r$ does not guarantee the existence of a solution for problem P2. In other words, given an arbitrary $K < N_r$ setup, although we have more receive antennas than interferers, there still may not be an analog cancellation vector that can cancel all the interferences from different SWIPT BSs due to the unit-modulus constraints. Responding to this, a sufficient condition on the maximum number of SWIPT BSs guaranteeing the feasibility of the formulated problem is analyzed  in the sequel.

\subsection{Summary of  Kronecker Based Construction}
Basically, the proposed Kronecker based construction is motivated by a key observation that each of the SWIPT channel vectors ${\bf v}(\Theta_i)$ possesses a special Vandermonde structure due to the free-space propagation environment, which enables it to be decomposed into a series of sub-vector components connected by Kronecker product (this decomposition is referred to as \emph{Kronecker decomposition}). Then, by exploiting the \emph{mixed-product property} of the Kronecker product, the multiple-SWIPT suppression constraints targeting at simultaneously nulling multiple SWIPT channels can be translated to several individual single-SWIPT suppression sub-constraints targeting at only one specific SWIPT channel, which can be easily handled by utilizing the Fourier based construction method proposed in the case of single SWIPT BS. In particular, the said Kronecker based construction approach has the procedure as summarized below and is elaborated in the sequel. 

\begin{itemize}

\item[1)] {\bf Offline Construction}

\begin{itemize}

\item {\bf Step  $\bf 1$} [Generation of a single cancellation vector]: A  cancellation vector ${\bf s}$ is designed to be a Kronecker product of a set of component phase-shift vectors, each component of which targets for suppressing only one specific SWIPT signal according to a predefined order, e.g., $O = \Theta_{1}\rightarrow \Theta_{2}\cdots \rightarrow \Theta_{{K}}$, called a \emph{SWIPT-cancellation order}. See details in sub-section~B.1).

 \item{\bf Step  $\bf 2$} [Generation of the cancellation vector mother set]: A mother set of cancellation vectors, denoted as $\mathcal{S}$, is generated using the same construction framework in Step $1$ with two specific techniques, called \emph{Fourier based cancellation} and \emph{cancellation order rearranging}. The constructed mother set consists of  a series of orthonormal subsets where the cancellation vectors are mutually orthogonal and have the same norm. See details in sub-section~B.2).

\item {\bf Step  $\bf 3$} [Selection of linearly independent cancellation vectors]:  $(N_r - K)$ linearly independent vectors are selected from the mother set, $\mathcal{S}$,  obtained in Step~$2$ using a simple greedy/random search algorithm, giving the desired solution for the optimization problem in (\ref{P:1}).  See details in sub-section~B.3). 
\end{itemize}

\item[2)] {\bf Adaptive Analog Spatial Cancellation}

The preceding procedure of offline construction constructs a set of linear functions $\{f_{m,n}\}$ mapping the parameters $\{\Theta_i\}$  to the phase shift elements of $\bS$. {Interestingly, the linear independence of the selected cancellation vectors in the step 3 of the above offline construction is observed from simulation to be almost invariant to the changes of the parameters $\{\Theta_i\}$.}\footnote{Extensive simulations show that the change of the parameters $\{\Theta_i\}$ just affects the condition number of the selected $\bS$, while the linear independence of the row vectors of $\bS$ still maintains as long as the values of $\{\Theta_i\}$ are distinct, i.e., ${\Theta_1}\neq{\Theta_2}\neq\cdots\neq{\Theta_K}$.} Consequently,
in the presence of mobility (time varying $\{\Theta_i\}$), adaptive analog spatial cancellation is simple and involves computing the cancellation matrix $\bS$ using $[\bS]_{m, n} = e^{f_{m, n}(\Theta_1, \cdots, \Theta_K)}$ and adjusting the phase shifts in the analog circuit accordingly.  The offline construction advantage of the proposed Kronecker based construction framework makes the analog spatial cancellation technique suitable for implementation at low-complexity mobile devices.

\end{itemize}

\begin{figure}[tt]
\centering
\includegraphics[width = 8cm]{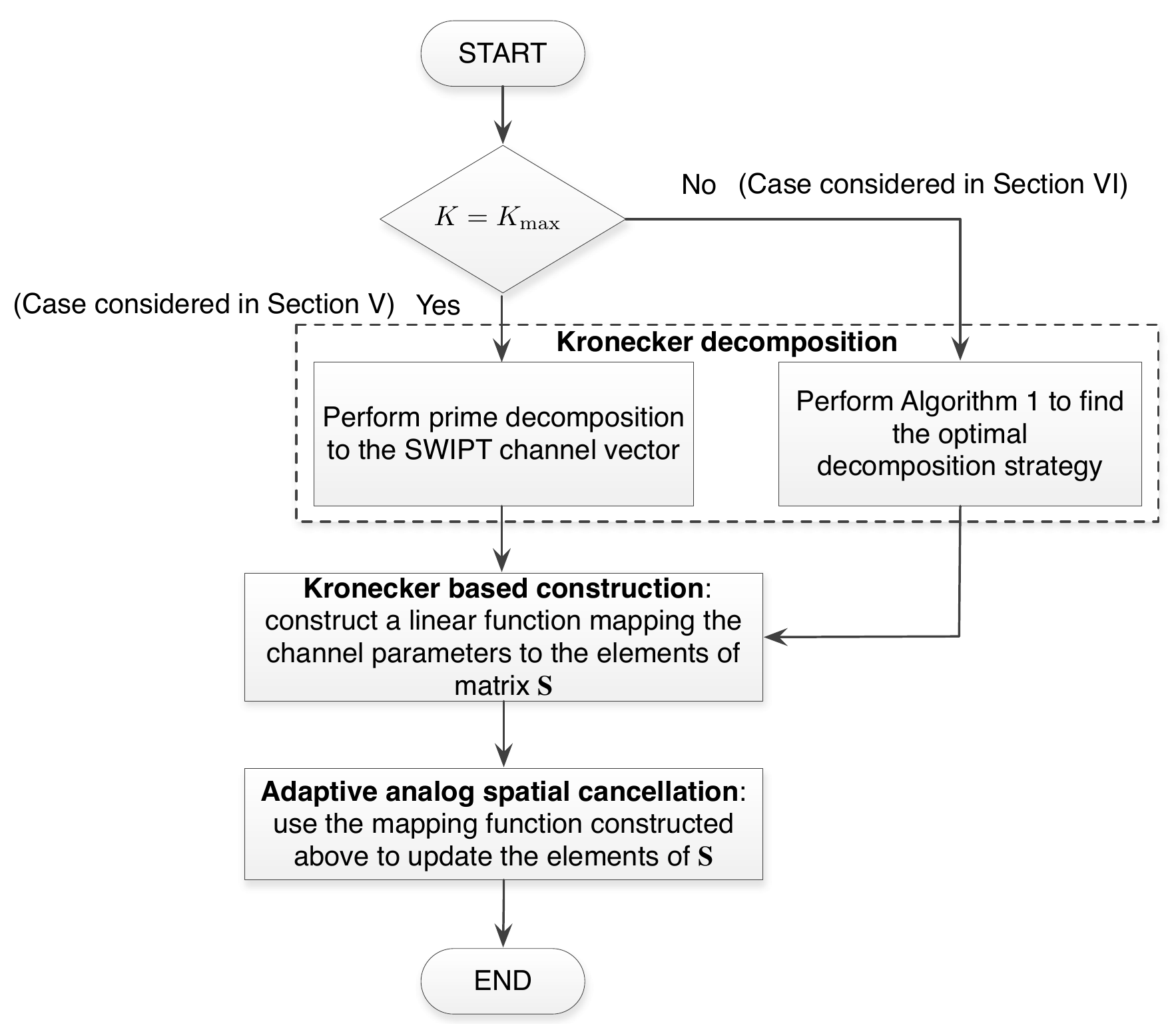}
\caption{ Flow chart illustrating the Kronecker based construction procedures for the case of multiple SWIPT BSs.}
\label{Fig:Flow_chart}
\end{figure}

An example illustrating the above procedure of Kronecker based construction is provided in Sub-section~C. 
Before we proceed to the discussion of the detailed procedure design, an unified flow chart showing the whole picture including both the $K=K_{\max}$ case in this section and the $K<K_{\max}$ case in the next section is provided in Fig. \ref{Fig:Flow_chart} to facilitate the reader to gain a better overview of the proposed Kronecker based construction.

\subsection{Detailed Design for Kronecker Based Construction}
\subsubsection{Generation of a single cancellation vector} \label{Step:1}
As mentioned earlier, the Vandermonde structure of the SWIPT channel vectors makes it possible to be decomposed into a Kronecker product of component phase-shift vectors. Moreover, it can be shown that the Kronecker decomposition of ${\bf v}(\Theta_i)$ is highly related to the factorization of its length $N_r$.
To be specific, we have the following key result showing the relationship between the factorization of $N_r$ and the Kronecker decomposition of ${\bf v}(\Theta_i)$.
\begin{proposition}\label{prop:3}
given $N_r = n_1n_2\cdots n_{K}$, with $n_1,n_2,\cdots,n_{K}$ being positive integers, the Kronecker decomposition of ${\bf v}(\Theta_i) $ can be given by
\begin{multline}\label{eq:p3}
{\bf v}(\Theta_i) = \\ [1,e^{j\Theta_i},\cdots,e^{j(n_1-1)\Theta_i}]\otimes [1,e^{jn_1\Theta_i},\cdots,e^{j(n_2-1)n_1\Theta_i}]\otimes \cdots \\
 \otimes[1,e^{jn_{K - 1}\cdots n_2n_1\Theta_i},\cdots,e^{j(n_{K} - 1)n_{K - 1}\cdots n_2n_1\Theta_i}],
\end{multline}
where $\otimes$ denotes the left Kronecker product operation.

\proof
It can be easily verified according to the definition of left Kronecker product \cite{HorJoh:MatrAnal:85}.
\endproof
\end{proposition}

\noindent Proposition \ref{prop:3} provides a general solution for performing Kronecker decomposition to a SWIPT channel vector given an arbitrary length $N_r$ and its factorization. The resultant Kronecker decomposition of the SWIPT channel vector is the most important step of the whole Kronecker based construction framework which motivates the construction of the corresponding cancellation vectors as specified in the sequel. 


Motivated by the Kronecker structure of the SWIPT channel vectors in Proposition~\ref{prop:3}, the desired cancellation vectors are designed to have the same structure as follows. 
\begin{figure}[t!]
\centering
\includegraphics[width=8cm]{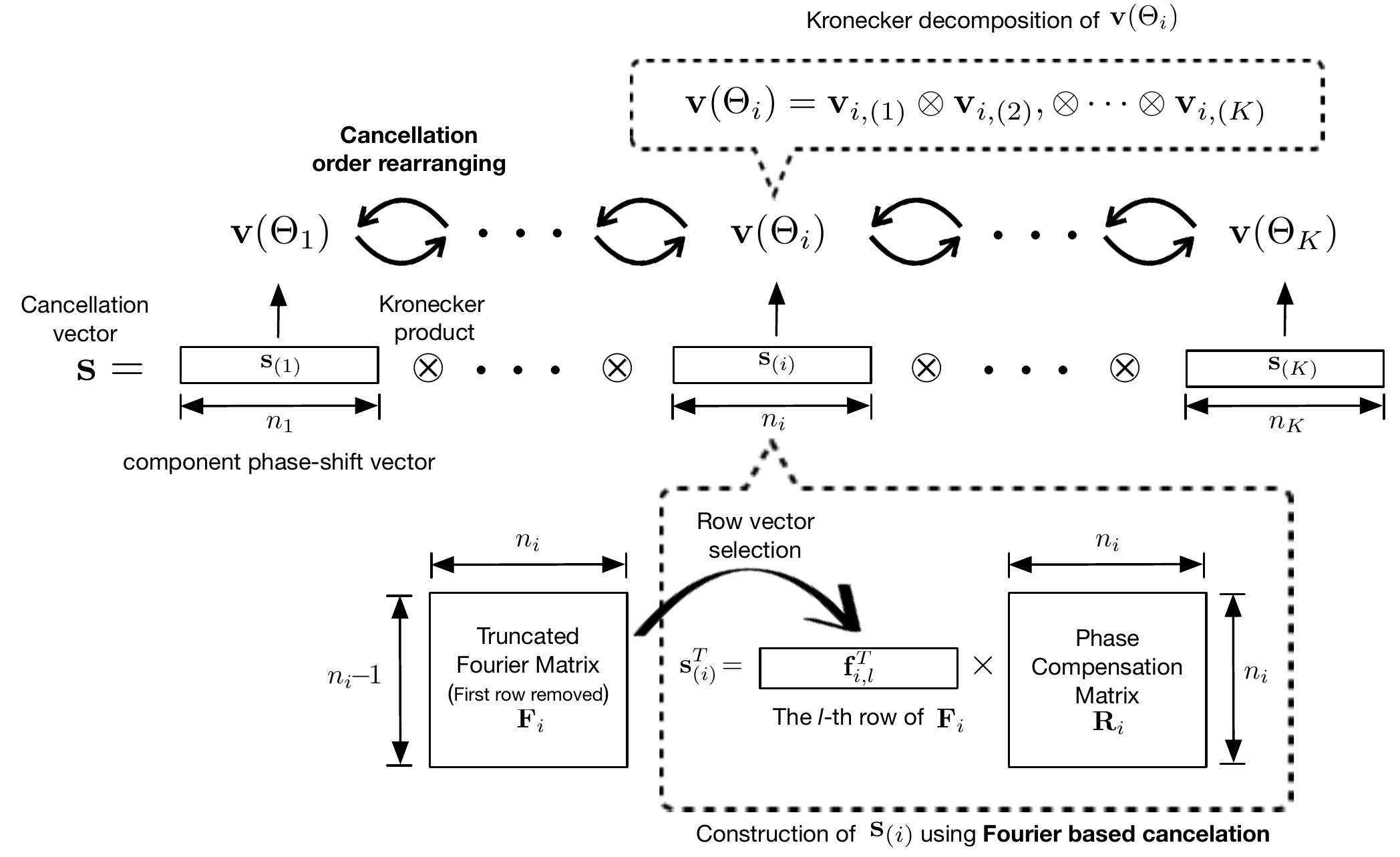}
\caption{Diagram illustrating the Kronecker based construction procedure.}
\label{Fig:Kronecker}
\end{figure}
For ease of  notation, write the Kronecker decomposition of ${\bf v}(\Theta_i)$ as  ${\bf v}(\Theta_i) = {\bf v}_{i,(1)} \otimes  {\bf v}_{i,(2)}, \otimes \cdots \otimes {\bf v}_{i,(K)}$ according to (\ref{eq:p3}), where ${\bf v}_{i,(j)}$ represents the $j$th component phase-shift vector of ${\bf v}(\Theta_i)$. Then, a typical cancellation vector, denoted by ${\bf s}$, can be constructed using the same Kronecker structure of ${\bf v}(\Theta_i)$ as shown in Fig. \ref{Fig:Kronecker}. Specifically, ${\bf s}$ consists of  $K$ component phase-shift vectors connected by Kronecker product, i.e., ${\bf s} =  {\bf s}_{(1)} \otimes {\bf s}_{(2)}\otimes \cdots \otimes{\bf s}_{(K)}$, where ${\bf s}_{(j)}$ denotes the $j$th component phase-shift vector of ${\bf s}$. For each component, the length of ${\bf s}_{(j)}$ is set to be the same as that of ${\bf v}_{i,(j)}$, denoted by $n_j$, satisfying $n_1n_2\cdots n_{K} = N_r$. Thus, the multiple-SWIPT suppression constraints can be rewritten as follows.
\begin{multline}\label{mp:0}
{\bf s}^T{\bf v}(\Theta_i) = {\bf s}_{(1)}^T{\bf v}_{i(1)}\otimes{\bf s}_{(2)}^T{\bf v}_{i(2)}\otimes \cdots \otimes {\bf s}_{(K)}^T{\bf v}_{i(K)} = 0,\\ \forall i\in\{1,2,\cdots,K\}.
\end{multline} 

Note that each one of the constraints in (\ref{mp:0}) related to a specific $i$ can be enforced if one of ${\bf s}_{(j)}^T{\bf v}_{i(j)} = 0$, $j\in\{1,2,\cdots,K\}$ holds. In other words, each component phase-shift vector ${\bf s}_{(j)}$ can be designed individually for tackling only one of the SWIPT channels among all total $K$ ones, then  Kronecker combining all the component phase-shift vector leads to a desired cancellation vector being orthogonal to all ${\bf v}(\Theta_i)$. Based on this key observation, the multiple-SWIPT suppression constraints can then be translated to $K$ single-SWIPT sub-constraints given a predefined SWIPT-cancellation order which indicates the mapping between the phase-shift-vector components and the targeted SWIPT channels. For example, given that the SWIPT-cancellation order is $O = \Theta_{1}\rightarrow \Theta_{2}\cdots \rightarrow \Theta_{{K}}$ (as depicted in Fig. \ref{Fig:Kronecker}), the $i$th component phase-shift vector of $\bs$ should be designed targeting at the $i$th SWIPT channel vector, then (\ref{mp:0}) can be translated to 
\begin{equation}\label{mp:1}
{\bf s}_{(i)}^T{\bf v}_{i(i)} = 0,\;\;\;\;\forall i\in\{1,2,\cdots,K\}.
\end{equation} 
For each single-SWIPT suppression sub-constraint in (\ref{mp:1}), it can be easily enforced by following the Fourier based construction method proposed in Section \ref{sec:5}. Specifically, a single component phase-shift vector  
${\bf s}_{(i)}^T$ can be constructed as a product of a Fourier row vector and a phase compensation matrix as shown in the following.
\begin{align}\label{mp:2}
 {\bf s}_{(i)}^T = {\bf f}_{i,l}^T{\bf R}_i,\;\;\; i\in\{1,2,\cdots,K\},
\end{align}
where ${\bf f}_{i,l}^T$ represents the $l$th row vector of a $(n_i - 1) \times n_i$ truncated Fourier matrix, denoted by ${\bf F}_i$, with $l$ being an arbitrary value among $l\in \{1,2,\cdots,n_i-1\}$; and ${\bf R}_i$ is the phase compensation matrix, which is a diagonal matrix with the diagonal elements being exactly the elements of ${\bf v}_{i(i)}^H$. To this end, by connecting all $K$ phase-shift-vector components via Kronecker product, 
the desired cancellation vector satisfying the multiple-SWIPT suppression constraints in (\ref {mp:0}) can be obtained. 

The above discussion also sheds light on the maximum number of cancellable SWIPT BSs that can be achieved by the proposed construction. 
To be specific, it can be noted that the number of cancellable SWIPT BSs is determined by the number of component phase-shift vectors that ${\bf v}(\Theta_i)$ can be decomposed to. In other words, $K_{\max}$ is achieved when the number of component phase-shift vectors in the Kronecker decomposition of ${\bf v}(\Theta_i)$ is maximized or, equivalently, when the number of factors in the facterization of $N_r$ is maximized. Note that prime decomposition gives the maximum number of factors as pointed out by the following  well known result from number theory. 

\begin{lemma}[Fundamental Theorem of Arithmetic]\label{lemma:1}
Every positive integer $N>1$ can be represented \textbf{in exactly one way} as a product of prime powers

\begin{equation}
N = p_1^{\alpha_1}p_2^{\alpha_2}\cdots p_t^{\alpha_t}
\end{equation}
where $p_1<p_2<p_3\cdots <p_t$ are primes and $\alpha_1,\alpha_2,\cdots,\alpha_t$ are positive integers, 
$t$ represents the maximum number of prime factors of integer $N$, and this representation is called the canonical representation.
\end{lemma}
Accordingly, $K_{\max}$ is obtained in the following proposition. 
\begin{proposition}[Maximum number of SWIPT BSs]\label{proposition:t1}
The  maximum number of SWIPT BSs that can be supported by analog spatial cancellation with $N_r$ receive antennas using the Kronecker based construction approach is given by
\begin{align}
K_{\max} = \alpha_1 + \alpha_2 + \cdots +\alpha_t,
\end{align}
where $\alpha_i$ denotes the $i$th power in the canonical representation of $N_r$ which is given as follows based on Lemma \ref{lemma:1}
\begin{align}\label{prime_decomposition}
N_r = p_1^{\alpha_1}p_2^{\alpha_2}\cdots p_t^{\alpha_t}.
\end{align}
\end{proposition}

\noindent Proposition \ref{proposition:t1} indicates the maximum number of cancellable SWIPT BSs, which is also served as a sufficient condition on the number of SWIPT BSs  ensuring the feasibility of problem P2 as mentioned eariler.
A close observation on this key result reveals that, from the perspective of achievable number of SWIPT BSs that can be supported, it is more ``efficient'' to set $N_r$ as an $n$th power of two, i.e., $N_r = 2^n$, $n \in \mathbb{Z}^+$. This is because that it requires the fewest number of receive antennas to perform analog spatial cancellation for a given $K$ compared with other values of $N_r$ comprising other prime factors larger than two. To this end, for the special case of $N_r = 2^n$, we have the following corollary.

\begin{corollary}
The  maximum number of SWIPT BSs that can be supported by analog spatial cancellation with $N_r = 2^n$ receive antennas is given as $K_{\max} = n$.
\end{corollary}


\subsubsection{Generation of the cancellation vector mother set}
Having obtained a single cancellation vector according to Step 1, next, Step 2 of the systematic procedure generates the mother set $\mathcal{S}$ of cancellation vectors based on the construction framework in Step 1 following the proposed Fourier based cancellation (FBC) and cancellation order rearranging (COR) approaches as follows.

Specifically, given a certain cancellation order $O$, FBC generates a set of cancellation vectors by selecting different Fourier row vectors to enforce the single-SWIPT suppression sub-constraints in (\ref{mp:1}). Note that there are $(n_i - 1)$ candidate Fourier row vectors that can be selected to achieve the construction of each component phase-shift vector ${\bf s}_{(i)}$ as indicated in (\ref{mp:2}), where $n_i$ denotes the length of vector ${\bf s}_{(i)}$ as shown in Fig. \ref{Fig:Kronecker}. Repeating Step 1 for all possible combinations yields $N_{\sf orth}$ cancellation vectors, where $N_{\sf orth}$ represents the number of the cancellation vectors provided by the FBC approach. Specifically, we have 
\begin{align}\label{mp:3}
N_{\sf orth}= (n_1 - 1)(n_2 - 1),\cdots,(n_{K} - 1),
\end{align}
where $n_1, n_2, \cdots, n_K$ are the factors of $N_r$ satisfying $N_r = n_1n_2\cdots n_K$.
Moreover, we have the following key property for the FBC construction approach.

%

%

\begin{proposition}\label{lemma:5}
All those cancellation vectors generated by the FBC construction approach following the same cancellation order are mutually orthogonal.

\begin{proof}
See Appendix \ref{appendix:lemma:5}.
\end{proof}
\end{proposition}


Note that the above construction involves only a single SWIPT-cancellation order. Actually, for each SWIPT-cancellation order, the same FBC approach can be repeated so that more cancellation vectors can be generated, which is the key idea of the COR approach. In particular, by repeating the construction for all $K!$ SWIPT-cancellation orders, the COR approach expands the vector set generated by the FBC approach to a $K!$ times larger mother set consisting of total $N_{\sf MS} = N_{\sf orth}K!$ candidate cancellation vectors. To this end, based on Proposition \ref{lemma:5}, we have the following key property for the proposed Kronecker based construction framework.


\begin{proposition}\label{prop:4}
Given $N_r = n_1n_2\cdots n_{K}$, the mother set $\mathcal{S}$ generated by the proposed Kronecker based construction procedure is composed of $N_{\sf sub} = K!$ subsets with each containing $N_{\sf orth} = (n_1-1)(n_2-1)\cdots (n_{K}-1)$ mutually orthogonal vectors.
\end{proposition}

Proposition \ref{prop:4} points out the partial orthogonal property of the constructed mother set, which can be further exploited to develop a more efficient greedy search algorithm in the sequel for finding the linearly independent cancellation vectors. 
Furthermore, with the help of the Lemma \ref{lemma:2} shown in the following, it can be proven that $N_r - K \leq N_{\sf MS}$. This reveals that the proposed systematic procedure can always generate a sufficiently large mother set with the number of cancellation vectors being no less than the required ones.

\begin{lemma}\label{lemma:2}
For any $2 \leq P_1\leq P_2 \leq \cdots \leq P_n $, $P_i \in \mathbb{Z}$, $i = 1,2,\cdots,n$, $n \in \mathbb{Z}^+$, the following inequality holds,
\begin{align}\label{SP:2}
P_1P_2\cdots P_n - n \leq (P_1 - 1)(P_2 - 2)\cdots (P_n - 1)n!.
\end{align}

\proof
See Appendix \ref{appendix:lemma:2}.
\endproof
\end{lemma}


\subsubsection{Selection of linearly independent cancellation vectors}\label{sec:greedy search}
 Last, a set of $(N_r - K)$ linearly independent vectors can be easily chosen from $\mathcal{S}$ using one of the well-known search methods such as exhaust search, greedy search and random search.  Specifically, exhaust search exhausts a complete search of all possible $(N_r - K)$ vectors combinations from the generated mother set to guarantee a global optimal solution in terms of condition number, which also leads to the highest implementation complexity. Greedy search does not pursue a global optimal solution, instead, it looks for a local optimal solution in each iteration (in our case, it is based on a criterion that the incremental matrix formed by the target vector and all those selected vectors should have the minimum condition number). Simulation results show that it can achieve almost the same performance of the exhaust search method with moderate complexity.
Last, random search adopts the simplest search strategy which randomly picks $(N_r - K)$ cancellation vectors from the mother set in each iteration and stops when the selected vectors are linearly independent. From simulations, it is observed that random search can always achieve a much lower complexity than the other two methods at the expense of moderate performance loss (in terms of condition number). Since mobile users are usually resource-constrained, implementation complexity should be given the highest priority when selecting a search algorithm. In this regard, random search method may be more preferred in mobiles from the practical perspective. Furthermore, it is worth pointing out that a subset of  $(N_r - K)$ linearly independent vectors can always be found out by any one of the three mentioned search methods, which also indicates that there always exists a desired linearly independent subset of the mother set.

\subsection{An Example}
To demonstrate the proposed systematic construction procedure, an example is provided below.

\begin{example}\label{example:1}
\emph{Let's consider the simple case with $N_r = 6$ and $K = 2$. According to Proposition \ref{proposition:t1}, it is noted that $K = 2 = K_{\max}$ and the corresponding prime factorization of $N_r$ is $N_r = 2\times3$, which gives the Kronecker decomposition of ${\bf v}(\Theta_1)$ and ${\bf v}(\Theta_2)$ based on Proposition \ref{prop:3} as follows.}
\begin{align}
{\bf v}(\Theta_i) = {\bv}_{i,(1)} \otimes {\bv}_{i,(2)} = [1,e^{j\Theta_i}]\otimes[1,e^{j2\Theta_i},e^{j4\Theta_i}],
\end{align}
where $i\in\{1,2\}$.

\emph{Thus, the first cancellation vector ${\bs}_1$ can be constructed as a product of two component phase-shift vectors, denoted by ${\bs}_{1(1)}$ and ${\bs}_{1(2)}$ respectively. Without loss of generality, assuming the SWIPT-cancellation order is $O = \Theta_{1}\rightarrow \Theta_{2}$, and using (\ref{mp:2}), ${\bs}_{1(1)}$ and ${\bs}_{1(2)}$ can be given by 
\begin{align}
&{\bs}_{1(1)} = [1,e^{j\pi}] \left[
  \begin{array}{cc}
  1 & 0 \\
   0 & e^{-j\Theta_1} 
  \end{array}
\right], \\
&{\bs}_{1(2)} =[1, e^{j\frac{2}{3}\pi},e^{j\frac{4}{3}\pi}] \left[
  \begin{array}{ccc}
  1 & 0 & 0 \\
   0 & e^{-j2\Theta_2} & 0\\
   0 & 0 &  e^{-j4\Theta_2}
  \end{array}
\right],
\end{align}
where ${\bs}_{1(2)}$ is constructed using the first row of a $2\times3$ truncated Fourier matrix. Then, it follows that}
\begin{multline}
{\bf s}_1^T = {\bs}_{1(1)} \otimes   {\bs}_{1(2)} = [1, -e^{-j\Theta_1}, e^{j(\frac{2}{3}\pi-2\Theta_2)}, \\
-e^{j(\frac{2}{3}\pi-2\Theta_2-\Theta_1)}, e^{j(\frac{4}{3}\pi-4\Theta_2)}, -e^{j(\frac{4}{3}\pi-4\Theta_2-\Theta_1)}],
\end{multline}

\emph{Alternatively, ${\bs}_{1(2)}$ can be constructed using the second row of a $2\times3$ truncated Fourier matrix, which gives
\begin{align}
{\bs}_{1(2)}' = [1, e^{j\frac{4}{3}\pi},e^{j\frac{2}{3}\pi}] \left[
  \begin{array}{ccc}
  1 & 0 & 0 \\
   0 & e^{-j2\Theta_2} & 0\\
   0 & 0 &  e^{-j4\Theta_2}
  \end{array}
\right].
\end{align}
It follows that ${\bf s}_2^T$ can be constructed as}
\begin{multline}
{\bf s}_2^T = {\bs}_{1(1)} \otimes {\bs}_{1(2)}'  =  [1, -e^{-j\Theta_1}, e^{j(\frac{4}{3}\pi-2\Theta_2)}, \\
 -e^{j(\frac{4}{3}\pi-2\Theta_2-\Theta_1)}, e^{j(\frac{2}{3}\pi-4\Theta_2)}, -e^{j(\frac{2}{3}\pi-4\Theta_2-\Theta_1)}].
\end{multline}

\emph{It is easy to verify that these two cancellation vectors are orthogonal, i.e., ${\bf s}_1^H{\bf s}_2 = 0$, which agrees with the result in Proposition \ref{lemma:5}. }

\emph{Next, by following another SWIPT-cancellation order 
$O^{'} = \Theta_{2}\rightarrow \Theta_{1}$, and repeating the above construction procedure, we can obtain the other two cancellation vectors as follows,}
\begin{multline}
{\bf s}_3 = [1, -e^{-j\Theta_2}, e^{j(\frac{2}{3}\pi-2\Theta_1)}, -e^{j(\frac{2}{3}\pi-2\Theta_1-\Theta_2)},
\\ e^{j(\frac{4}{3}\pi-4\Theta_1)}, -e^{j(\frac{4}{3}\pi-4\Theta_1-\Theta_2)}]^T,
\end{multline}
\begin{multline}
{\bf s}_4 = [1, -e^{-j\Theta_2}, e^{j(\frac{4}{3}\pi-2\Theta_1)}, -e^{j(\frac{4}{3}\pi-2\Theta_1-\Theta_2)}, \\ e^{j(\frac{2}{3}\pi-4\Theta_1)}, -e^{j(\frac{2}{3}\pi-4\Theta_1-\Theta_2)}]^T.
\end{multline}

\emph{To this end, combining all the generated cancellation vectors yields the desired cancellation matrix as shown in (\ref{S:62}).}
\begin{figure*}
\begin{align}\label{S:62}
{\bf S}_{6,2} = [{\bf s}_1, {\bf s}_2, {\bf s}_3, {\bf s}_4]^T =
\left[
  \begin{array}{cccccc}
  1 & -e^{-j\Theta_1} & e^{j(\frac{2}{3}\pi-2\Theta_2)} & -e^{j(\frac{2}{3}\pi-2\Theta_2-\Theta_1)} & e^{j(\frac{4}{3}\pi-4\Theta_2)} & -e^{j(\frac{4}{3}\pi-4\Theta_2-\Theta_1)}\\
  1 & -e^{-j\Theta_1} & e^{j(\frac{4}{3}\pi-2\Theta_2)} & -e^{j(\frac{4}{3}\pi-2\Theta_2-\Theta_1)} & e^{j(\frac{2}{3}\pi-4\Theta_2)} & -e^{j(\frac{2}{3}\pi-4\Theta_2-\Theta_1)} \\
  1 & -e^{-j\Theta_2} & e^{j(\frac{2}{3}\pi-2\Theta_1)} & -e^{j(\frac{2}{3}\pi-2\Theta_1-\Theta_2)} & e^{j(\frac{4}{3}\pi-4\Theta_1)} & -e^{j(\frac{4}{3}\pi-4\Theta_1-\Theta_2)} \\
  1 & -e^{-j\Theta_2} & e^{j(\frac{4}{3}\pi-2\Theta_1)} & -e^{j(\frac{4}{3}\pi-2\Theta_1-\Theta_2)} & e^{j(\frac{2}{3}\pi-4\Theta_1)} &  -e^{j(\frac{2}{3}\pi-4\Theta_1-\Theta_2)}\\
  \end{array}
\right].
\end{align}
\hrule
\end{figure*}

\end{example}
It can be easily verified that ${\bf S}_{6,2}$ is the desired solution of the problem in (\ref{P:2}) for the case of $N_r = 6$, since ${\sf Rank} ({\bf S}) = N_r - K = 4$, given $\Theta_1 \neq \Theta_2$, and the unit-modulus constraints are also perfectly satisfied.
\begin{Remark}
 \emph{In this simple example, the size of mother set $\mathcal{S} = \{\bs_1,\bs_2,\bs_3,\bs_4\}$ is equal to that of the final cancellation vector set, i.e., $N_{\sf MS}= N_r - K$ and the linear independency of the generated vectors can be easily verified. 
 For the more complex case where $N_{\sf MS} > N_r - K$, it can be shown that the required $(N_r - K)$ linear independent vectors can still be selected from $\mathcal{S}$ by using the said greedy or random search algorithm (see the example in Fig. \ref{Fig:independent set}). }
 \end{Remark}

Last, adapting  analog spatial cancellation to time-varying $(\Theta_1, \Theta_2)$ involves re-computing ${\bf S}_{6,2} $ using \eqref{S:62} upon changes on  the parameters.




\section{Analog Spatial Cancellation for The Case of $K < K_{\max}$}\label{sec:7}
We now look into the case that $K < K_{\max}$. The main difference between this case and the prior case is that the less number of SWIPT signals that needs to be analog decoupled leads to an non-unique Kronecker decomposition strategy of ${\bf v}(\Theta_i)$ or, equivalently, an non-unique factorization strategy of $N_r$. For example, given $N_r = 12$ and $K = 2$, we have two factorization strategies for tackling this case, i.e., $12 = 2\times 6$ and $12 = 3\times 4$, which respectively correspond to two Kronecker decomposition strategies of ${\bf v}(\Theta_i)$ as follows,
\begin{align}
&{\bf v}(\Theta_i) = [1, e^{j\Theta_i}]\otimes[1,e^{j2\Theta_i},e^{j4\Theta_i},e^{j6\Theta_i},e^{j8\Theta_i},e^{j10\Theta_i}].\label{mpb:1} \\
&{\bf v}(\Theta_i) = [1, e^{j\Theta_i},e^{j2\Theta_i}]\otimes[1,e^{j3\Theta_i},e^{j6\Theta_i},e^{j9\Theta_i}].\label{mpb:2}
\end{align}
Note that (\ref{mpb:1}) and (\ref{mpb:2}) correspond to different cancellation vector construction processes as indicated in Section \ref{sec:6}, which results in different mother set with different size and $N_{\sf orth}$.  This raises additional design problem for selecting the proper factorization strategy in this case. To be specific, (\ref{mpb:1}) leads to a mother set containing $10$ cancellation vectors with $N_{\sf orth} = 5$, while (\ref{mpb:2}) results in another mother set comprising $12$ vectors with $N_{\sf orth} = 6$. Apparently, the linearly independent subset of cancellation vectors selected form the latter would achieve a lower condition number than that of those selected from the former due to the fact that the latter is larger mother set with more mutually orthogonal cancellation vectors. In this regard, it is more desired to choose a factorization strategy leading to the largest $N_{\sf orth}$.
Mathematically, assuming the factorization of $N_r$ can be expressed as $N_r = n_1n_2\cdots n_{K}$, the optimal factorization strategy selection problem, given a certain $N_r$ and $K$, can be formulated as follows.

\begin{equation}
\begin{aligned}
\mathop {\max }\limits_{n_1,n_2,\cdots,n_{K}} \; &(n_1 - 1)(n_2 -1)\cdots (n_{K} -1) \\
{\textmd{s.t.}}\;\;\;\;& n_1n_2\cdots n_{K} = N_r \\
& n_1,n_2,\cdots,n_{K} \in \mathbb{Z}^+ .
\end{aligned}
\label{P:5}
\end{equation}

Note that it is a challenging integer programming problem with polynomial objective function and constraint. To tackle this problem, we have the following Lemma.
\begin{lemma}\label{lemma:4}
For any $2 \leq P_1\leq P_2 \leq \cdots \leq P_m $, $P_i \in \mathbb{Z}^+$, $i = 1,2,\cdots,m$, $m \in \mathbb{Z}^+$, the following inequality holds,
\begin{align}\label{NPB:2}
(P_1-1)(P_2-1)\cdots (P_m -1) \leq (\sqrt[m]{P_1P_2\cdots P_m}-1)^m.
\end{align}
where the equality holds if and only if $P_1=P_2=\cdots=P_m$.

\proof
See Appendix \ref{appendix:lemma:4}.
\endproof
\end{lemma}

Lemma \ref{lemma:4} reveals that the possible maximum value of the objective function in (\ref{P:5}) is
\begin{align}
(\sqrt[K]{N_r}-1)^{K},
\end{align}
and it is achieved only when $N_r$ is evenly factorized, i.e., $n_1 = n_2 = \cdots = n_{K}$. However, this solution may not lie in the feasible set of the problem in (\ref{P:5}) due to the integer factors constraints. Alternatively, the optimal solution for the problem in (\ref{P:5}) requires all factors of $N_r$ to be as even as possible. To this end, the following Algorithm \ref{algorithm:2} is proposed to attain the optimal factorization solution for the problem (\ref{P:5}).

\begin{algorithm}[ht]
\textbf{Input}:

$N_r$ : the number of receive antenna;

$K$ : the number of SWIPT BS;

\textbf{Output}:

$n_1,n_2,\cdots,n_{K}$ : the optimal factorization strategy;

1: Prime factorize $N_r$ according to (\ref{prime_decomposition})

2: Select the two smallest factors in (\ref{prime_decomposition}) and combine them as a composite, then the achievable
  $K$ reduces to $K_{\max} - 1$, and we obtain a new factorization expression

3: \textbf{repeat}

4: Select the two smallest factors in the new factorization expression and combine them as a composite to generate a new factorization expression. The achievable $K$ minus one after each iteration.

5: \textbf{until} achievable $K$ reduces to the target SWIPT BS number.
\caption{Iterative Algorithm for finding the optimal factorization strategy}\label{algorithm:2}
\end{algorithm}

Having obtained the optimal factorization solution for problem in (\ref{P:5}), we can design the optimal Kronecker decomposition strategy according to Proposition \ref{prop:3}, and then following the Kronecker based construction framework proposed in Section \ref{sec:6}, a corresponding mother set $\mathcal{S}$ can be generated. Finally, by employing one of the search algorithms mentioned in Section \ref{sec:greedy search}, we can select the desired $(N_r - K)$ linearly independent cancellation vectors from the mother set to form a full row rank phase shift matrix as a solution for the problem P2. 

Furthermore, the above discussion also implies an interesting tradeoff between $K$ and $N_{\sf orth}$ as expressed in the following proposition.

\begin{proposition} {\label{prop:7}}
Given a certain $N_r$, a smaller $K$ leads to a larger $N_{\sf orth}$. 

\proof
Without loss of generality, assume that the optimal factorization strategy of $N_r$ for a given $K$ is
$N_r = n_1n_2\cdots n_K$, where $2 \leq n_1\leq n_2 \leq \cdots \leq n_K$. Then, using (\ref{mp:3}), the corresponding $N_{\sf orth}$ can be computed as follows
\begin{align}
N_{\sf orth}^{(K)} = (n_1 -1)(n_2 - 1)\cdots(n_K - 1).
\end{align}

Next, consider the case that the number of SWIPT BS is $K-1$. According to Algorithm \ref{algorithm:2}, 
the optimal factorization strategy can be expressed as $N_r = m_1m_2\cdots m_{K-1}$, where $m_1 = n_1n_2$ and $m_i = n_{i+1}$, $i = 2,3,\cdots K-1$.  Then we have
\begin{align}
N_{\sf orth}^{(K-1)} &= (m_1 -1)(m_2 - 1)\cdots(m_{K-1} - 1) \notag\\
&= (n_1n_2-1)(n_3 -1)\cdots (n_K-1).
\end{align}

Note that $\frac{N_{\sf orth}^{(K)} }{N_{\sf orth}^{(K-1)} } = \frac{(n_1-1)(n_2-1)}{n_1n_2-1} < 1$, since $n_1,n_2 \geq 2$. To this end, the desired result can be proven by induction.
\endproof
\end{proposition}


{
\begin{Remark}\label{remark:prop:7}
\emph{ Proposition \ref{prop:7} characterizes the tradeoff between $K$ and $N_{\sf orth}$, which also implicitly reflects the relationship between $K$ and the achievable condition number of the constructed cancellation matrix ${\bf S}$.  In general, given a certain $N_r$, a larger number of SWIPT BSs $K$ leads to a smaller $N_{\sf orth}$, which in turn results in a degradation of the condition number performance. Simulation result will be provided in Fig. \ref{Fig:impact of K} to verify this analytical result.}
\end{Remark}
}


\section{Simulation Results}

\begin{figure}[tt]
\centering
\includegraphics[width = 8cm]{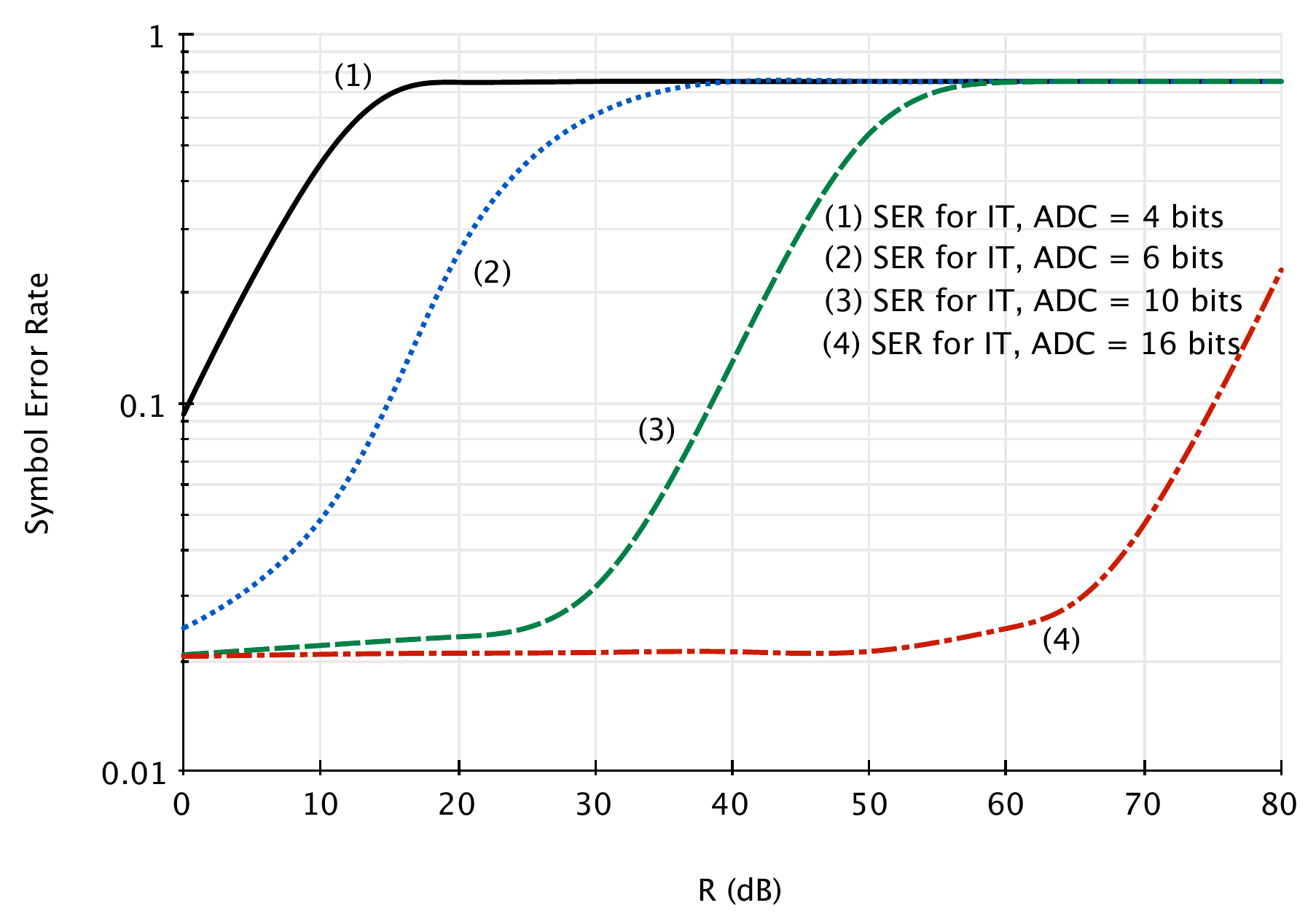}
\caption{ The SER performance of the traditional digital spatial cancellation for a varying SWIPT-IT signal power ratio.}
\label{Fig:ADC impact}
\end{figure}

In this section, the effectiveness of the proposed Kronecker based construction framework for finding the desired full row rank analog spatial cancellation matrix is first examined. Then, simulation results for further evaluating the performance of analog spatial cancellation are presented. The simulation settings for Fig.\ref{Fig:ADC impact}--\ref{Fig:imperfect cancellation} are summarized as follows. We set $N_r = 4$, $N_t = 4$, the average received SNR for the IT signal as $10\;{\sf dB}$. Moreover, the number of data streams for IT is fixed at $2$ for Fig. \ref{Fig:ADC impact}--\ref{Fig:SER comparison multiple PB case} and Fig. \ref{Fig:imperfect cancellation}, and fixed at $3$ for Fig. \ref{Fig:throughput comparison single PB case} and \ref{Fig:throughput comparison multiple PB case}, and that for SWIPT is $1$ per SWIPT BS, all modulated using QAM. It is also assumed that all the SWIPT signals are intended for the considered user, and the receiver at the user decouples the received mixed signals using the proposed analog spatial cancellation technique and decodes the SWIPT and IT signals separately.  { Perfect analog spatial cancellation is assumed for the results  in Fig. \ref{Fig:SER comparison single PB case}--\ref{Fig:throughput comparison multiple PB case}, and the impact of imperfect cancellation is investigated in Fig. \ref{Fig:imperfect cancellation}.  Finally, the impact of the number of SWIPT BSs on the condition number of the constructed cancellation matrix is illustrated in Fig. \ref{Fig:impact of K}.}

\begin{figure}[tt]
  \centering
  \subfigure[Exhaust Search]{\label{fig:4a}\includegraphics[width=0.4\textwidth]{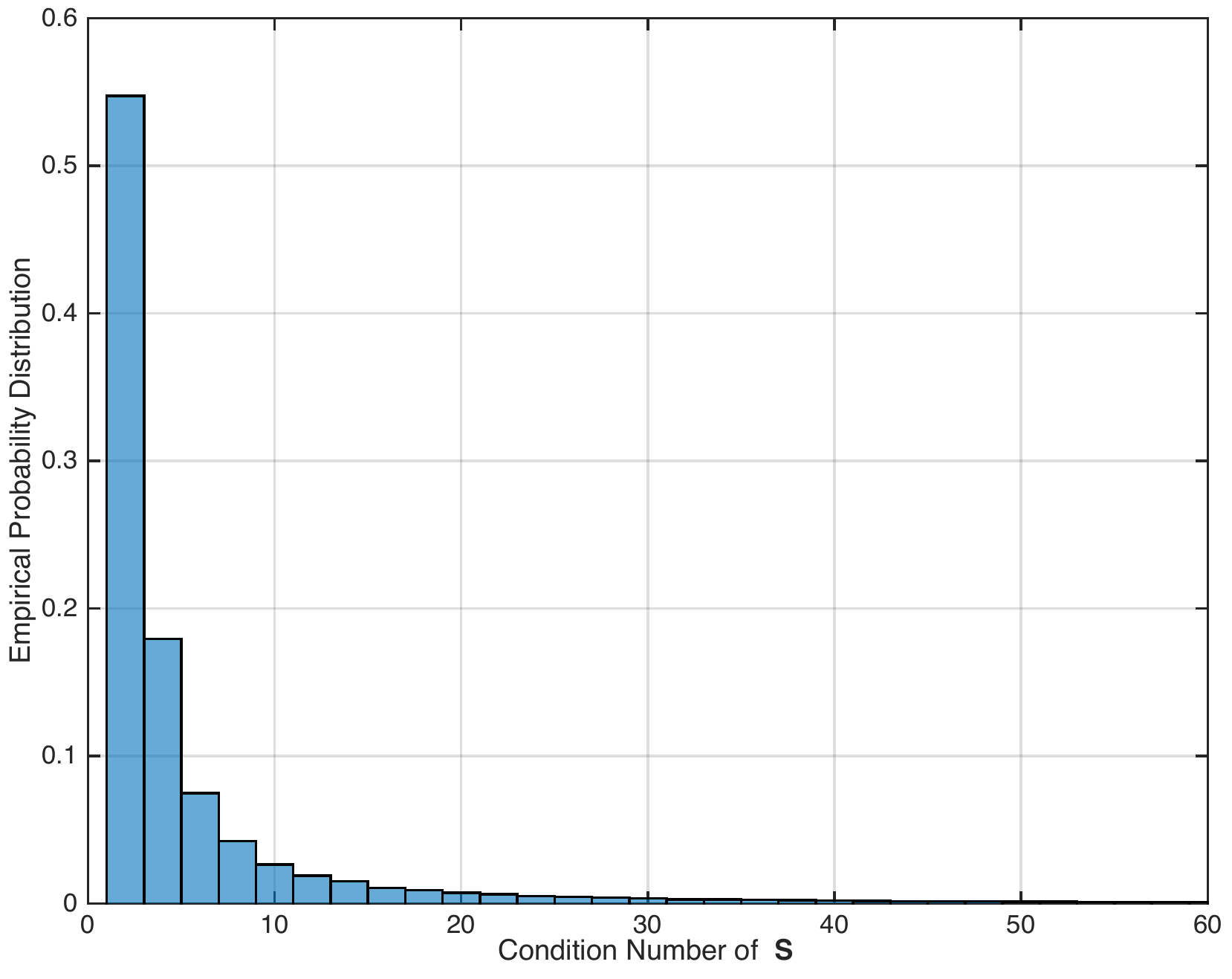}}
  \hspace{0.2in}
  \subfigure[Greedy Search]{\label{fig:4b}\includegraphics[width=0.4\textwidth]{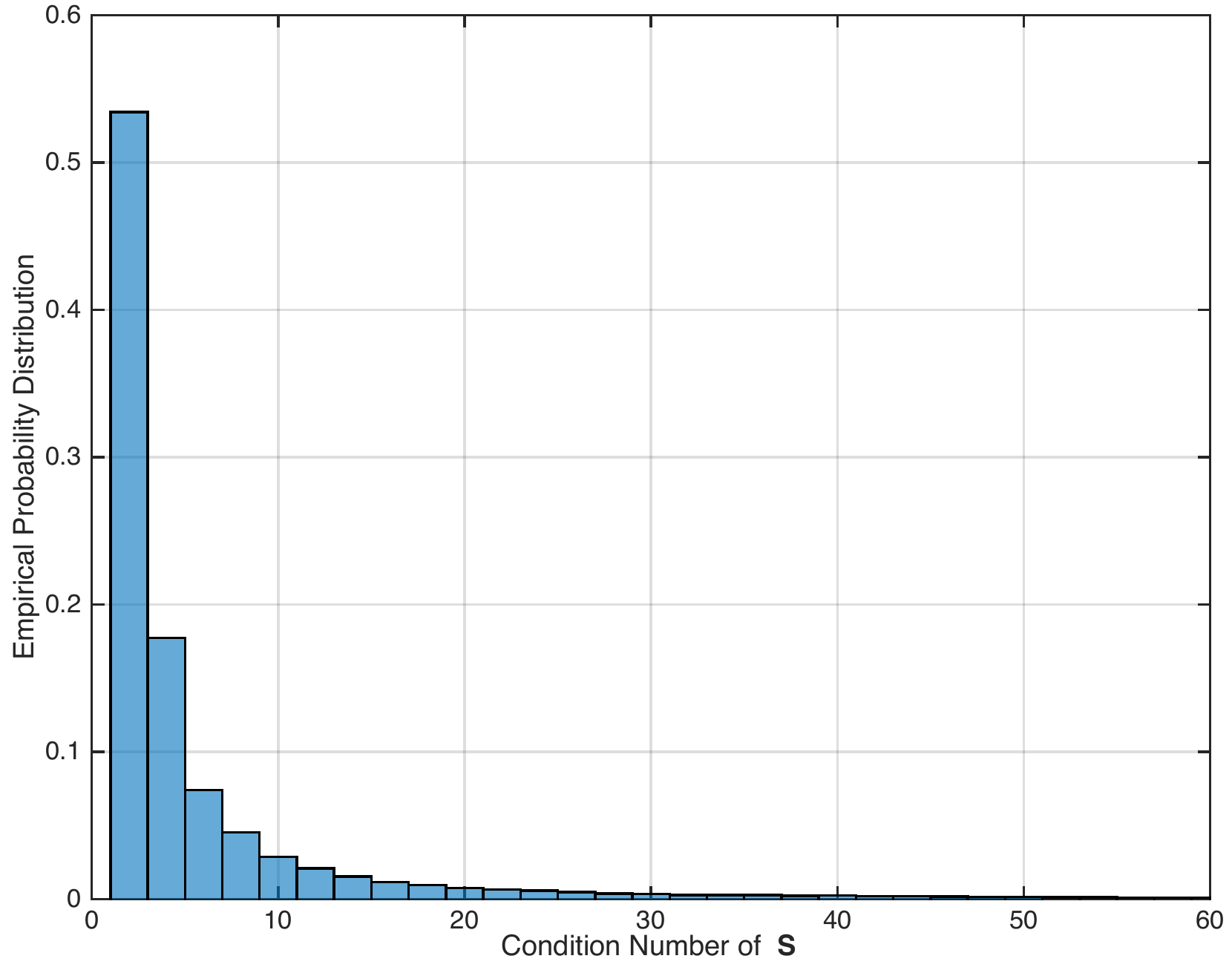}}
  \hspace{0.2in}
  \subfigure[Random Search]{\label{fig:4c}\includegraphics[width=0.4\textwidth]{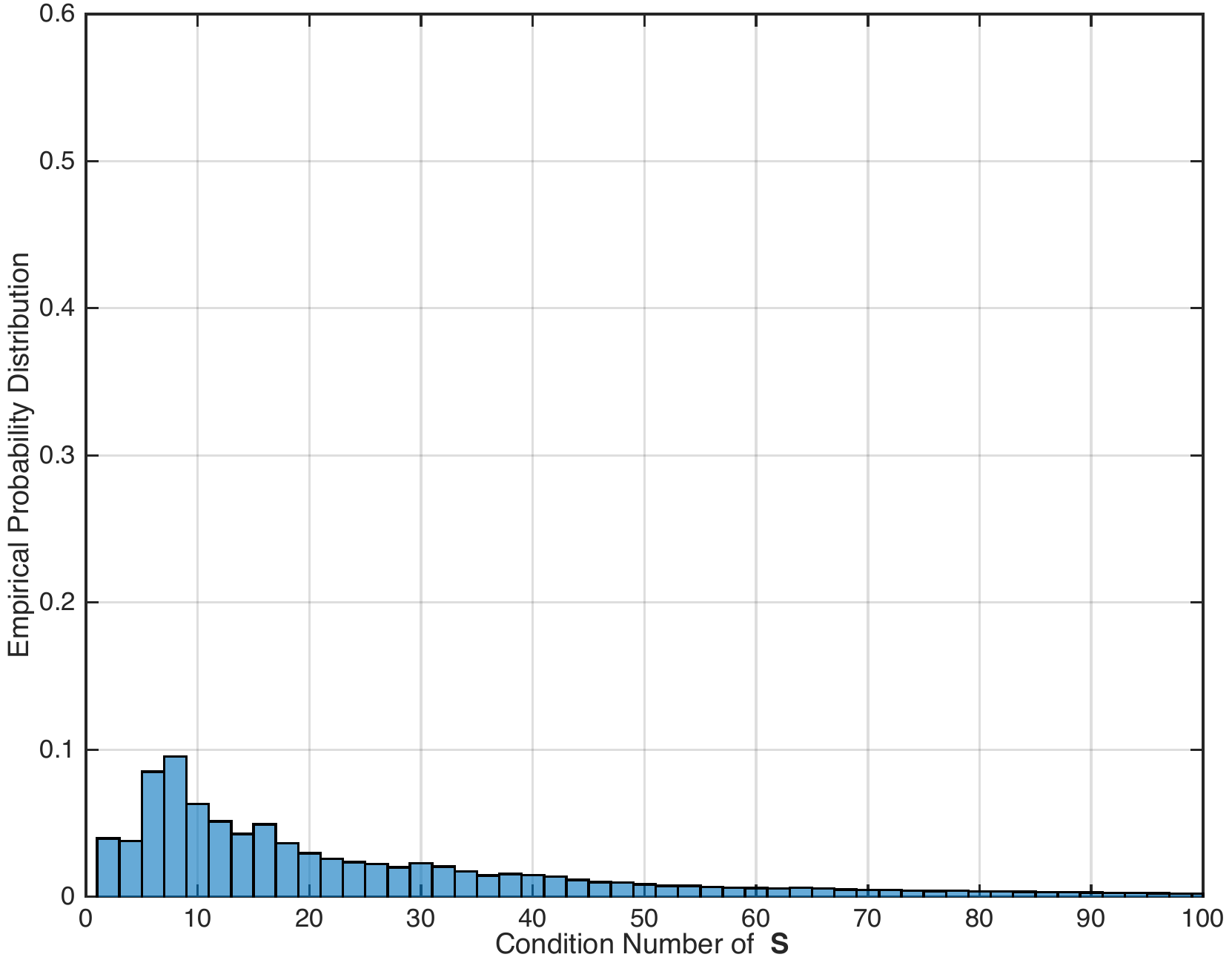}}
  \caption{Condition number distribution of the selected analog spatial cancellation matrix $\bf S$ for the case of $N_r = 12$ and $K = 2$, according to $10^6$ realizations.}
  \label{Fig:independent set}
\end{figure}

\subsection{Condition Number of Cancellation Matrix}
Considering the case of $N_r = 12$ and $K = 2$ with $\Theta_1$ and $\Theta_2$ uniformly distributed within $[0,2\pi]$, the { empirical probability distributions} of the condition number of the constructed cancellation matrix ${\bf S}$ obtained by the three mentioned search algorithms are plotted in Fig. \ref{Fig:independent set} respectively based on $10^6$ realizations.
It is observed that the selected ${\bf S}$ by the exhaust search and greedy search can achieve nearly minimum condition number (e.g., close to $1$) for most realizations, and the probability decays sharply as the condition number increases, leaving only negligible portion of the realizations resulting in a condition number larger than a moderate value (e.g., $30$). 
It is also noted that, all realizations lead to finite condition numbers which verifies the claim that there always exists a set of ($N_r - K$) linearly independent vectors in the generated mother set as mentioned in Section \ref{sec:greedy search}.
 However, the iterations required by the exhaust search and greedy search in the considered case are $66$ and $18$ respectively, which may be too computation-consuming to be implemented in some resource-limited mobiles. Alternatively, random search algorithm provides us with a low complexity solution requiring only about $1.1$ iterations in average, which implies that in most cases, the initial random selection is enough for finding a linearly independent subset solution and no further iteration is required. However, the low complexity is achieved at the expense of relatively bad condition numbers distribution as shown in Fig. \ref{fig:4c}, where the distribution looks quite even with a relatively smooth decay rate. Nevertheless, it is still an effective solution since the linear independence of the selected vectors can be guaranteed for all realizations of $\Theta_1$ and $\Theta_2$. 

\subsection{The PT-IT Near-Far Problem in Digital Spatial Cancellation}
Consider the use of the traditional digital spatial cancellation of IT and SWIPT signals in the presence of a single SWIPT BS. Fig. \ref{Fig:ADC impact} shows the symbol error rate (SER) for IT transmission versus the power ratio between the received SWIPT and IT signals, denoted as $R$. The SER is observed to depend on both the ADC resolution and $R$. Specifically, given a required SER, increasing $R$ (corresponding to a more severe near-far problem) requires higher ADC resolution so as to regulate the quantization noise in the weak IT signal. For high $R$ (e.g., $70\;{\sf dB}$ or higher), even a $16$-bit ADC is insufficient for achieving a low SER (e.g., 0.01). The results show that digital spatial cancellation is incapable of coping with the near-far problem, for which has to rely on using an ADC with an impractically high resolution.

\subsection{SER Comparison between Analog and Digital  Spatial Cancellation}
\begin{figure}[tt]
\centering
\includegraphics[width = 8cm]{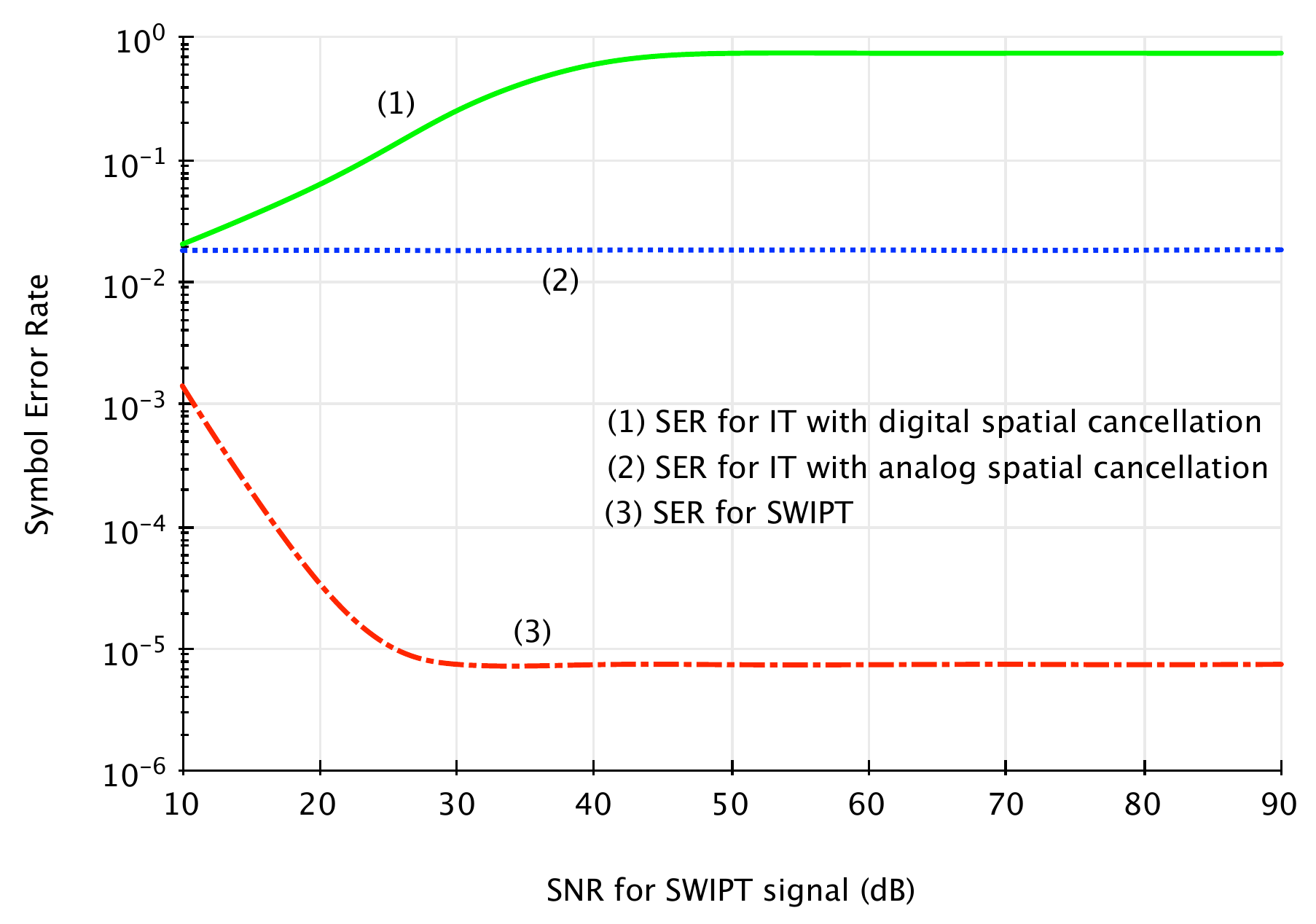}
\caption{SER comparison between the digital and analog spatial cancellation for the case of single SWIPT BS.}
\label{Fig:SER comparison single PB case}
\end{figure}

Fig. \ref{Fig:SER comparison single PB case}  compares the (IT transmission) SER performance of the proposed analog spatial cancellation and the traditional digital spatial cancellation in the case of single SWIPT BS. The channel noise variance and the average received SNR for the IT signal are assumed fixed and the ADC resolution is $6$ bits. It is observed that, for the digital cancellation, as the received SNR for the SWIPT signal increases, the SER for the IT signal grows and saturates at the worst point, i.e., $0.75$, which agrees with the SQNR analysis in (\ref{quan_error}) where the quantization noise for the IT signal is shown to be proportional to $R$. In contrast, when the proposed analog spatial cancellation is used, the (IT) SER performance is observed to be independent of the received SWIPT signal and attain a constant low SER throughout the whole SWIPT SNR range with only a $6$-bit ADC. This demonstrates the robustness of analog spatial cancellation against the near-far problem. In addition, it is also noted that the SER curve for the SWIPT signal has a floor due to quantization noise independent of the received SNR and determined only by the ADC resolution. 

\begin{figure}[tt]
\centering
\includegraphics[width = 8cm]{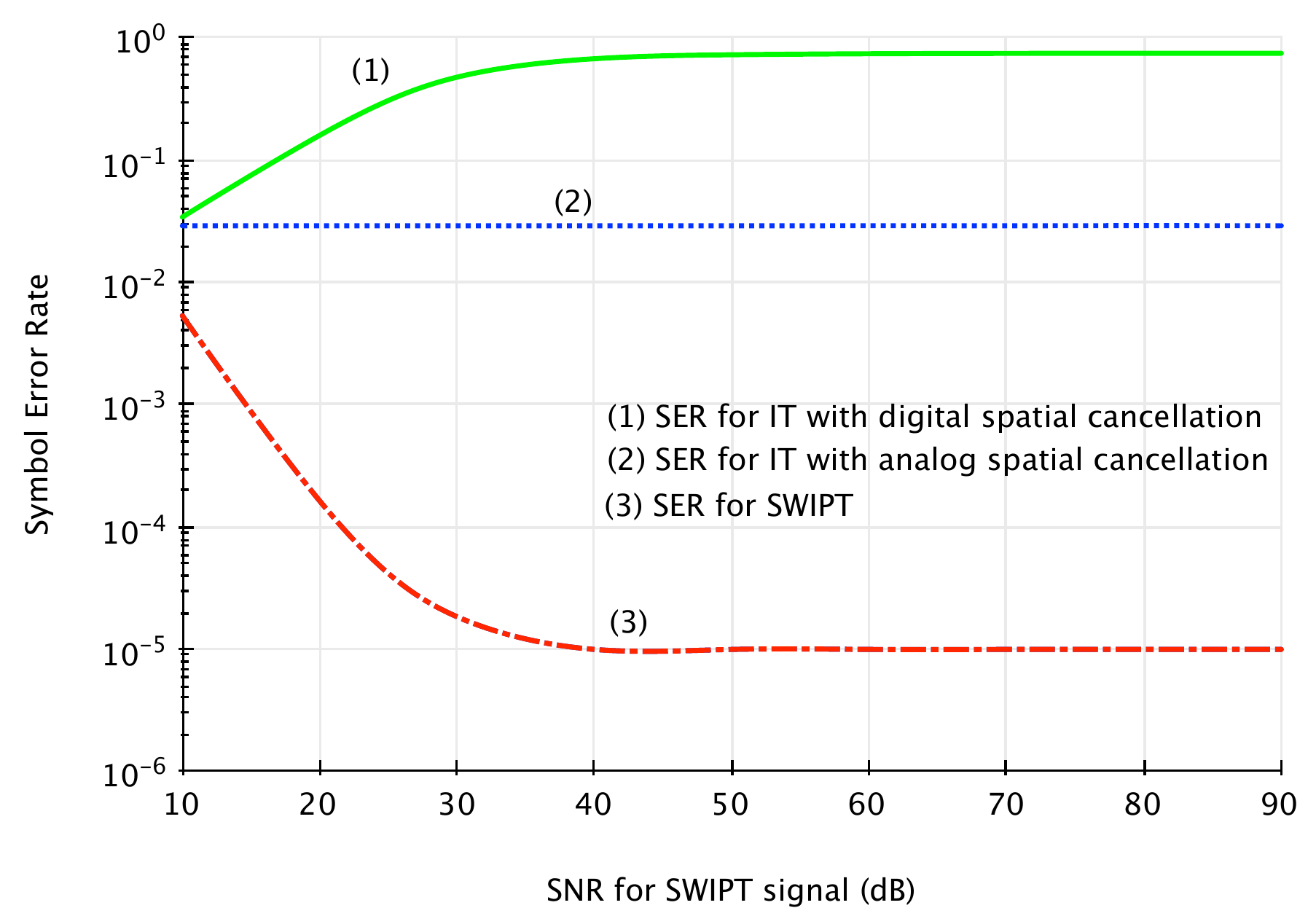}
\caption{SER comparison between the digital and analog spatial cancellation for the case of $K=2$.}
\label{Fig:SER comparison multiple PB case}
\end{figure}

SER comparison in the case of multiple SWIPT BS is illustrated in Fig. \ref{Fig:SER comparison multiple PB case}, where the system setup with two SWIPT BSs transmitting two independent SWIPT data streams is assumed, and the ADC resolution is also $6$ bits. As expected, similar trends can be observed in this case, which verifies the effectiveness of the proposed systematic solution for analog cancellation matrix in the case of multiple SWIPT BSs. Also, it is observed that, in terms of SER performance, single-SWIPT-BS case can achieve a better IT performance than the multiple-SWIPT-BS case given the number of supported IT data streams is the same. It is intuitive since additional DoF will be used to suppress the SWIPT signals for IT signals decoding in the case of multiple SWIPT BSs.

\subsection{Throughput Comparison between Analog and Digital  Spatial Cancellation}
\begin{figure}[tt]
\centering
\includegraphics[width = 8cm]{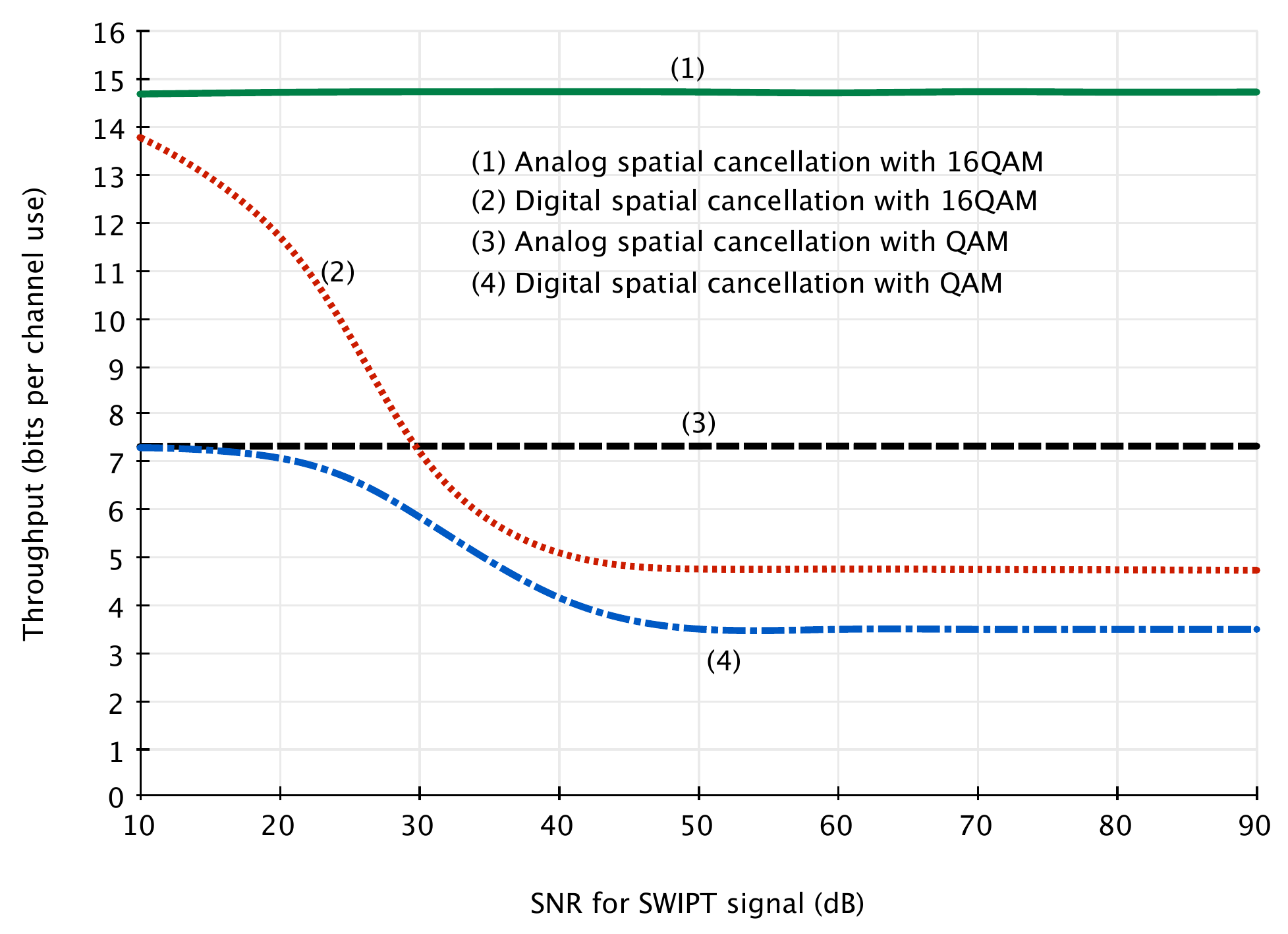}
\caption{Throughput comparison between the digital and analog spatial cancellation for the case of single SWIPT BS.}
\label{Fig:throughput comparison single PB case}
\end{figure}

\begin{figure}[tt]
\centering
\includegraphics[width = 8cm]{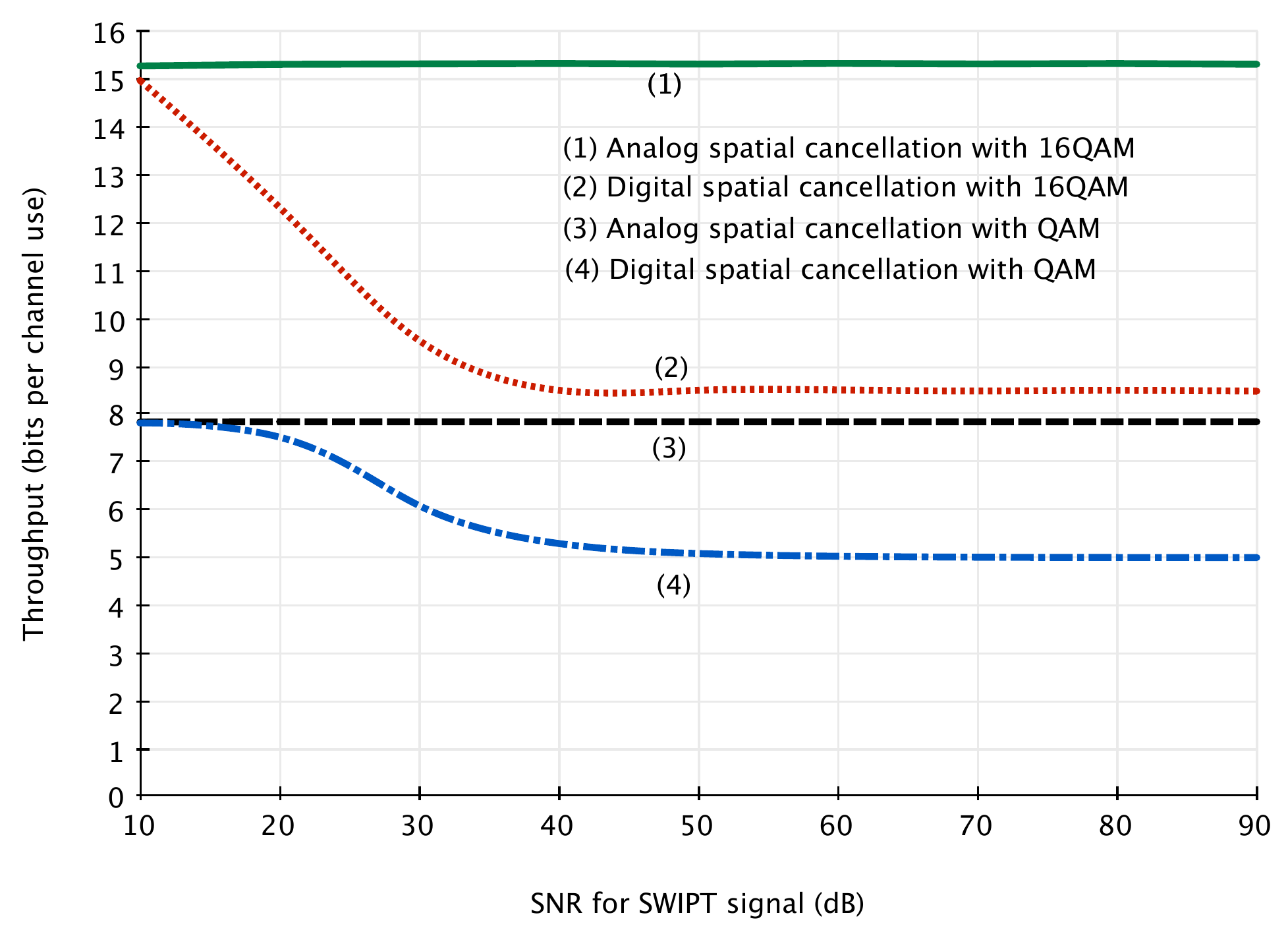}
\caption{Throughput comparison between the digital and analog spatial cancellation for the case of $K=2$.}
\label{Fig:throughput comparison multiple PB case}
\end{figure}

The throughput comparison between the proposed analog spatial cancellation and the traditional digital spatial cancellation for the case of single SWIPT BS is illustrated in Fig. \ref{Fig:throughput comparison single PB case}. 
 We consider the effective throughput in this comparison, which is defined by $\tau = (1 - P_{\sf SER}) \log_2M$, where $P_{\sf SER}$ is the symbol error rate and $M$ is the modulation order. 
 Several observations can be made. First, the throughput for the digital cancellation case decreases as the SWIPT SNR increases and saturates at a fixed point, while the throughput for the proposed analog spatial cancellation is unaffected by the strength of the SWIPT signal, which demonstrates the effectiveness of the analog spatial cancellation in tacking the near-far problem. Next, increasing the modulation order leads to a significant throughput improvement for the analog spatial cancellation case but only marginal performance increase for the digital counterpart. The superiority of the analog cancellation method is more obvious in the large SWIPT SNR regime (e.g., 
 $\rho_{\sf SWIPT} > 50\;{\sf dB}$), where the throughput of the analog cancellation case is observed to be nearly three times as that of the digital cancellation case given the modulation is 16QAM. 

Fig. \ref{Fig:throughput comparison multiple PB case} shows the throughput comparison in the case of multiple SWIPT BSs, where the system setup consists of two SWIPT BSs with two independent SWIPT streams is assumed. Besides those similar observations as shown in Fig. \ref{Fig:throughput comparison single PB case}. It is noted that, given the same MIMO configuration, the throughput performance of the case of multiple SWIPT BSs outperforms that of the case of single SWIPT BS, which is opposite to the trend observed in the previous SER comparison. This is due to the fact that the short-range SWIPT signal enjoys a much better channel condition than the long-range IT signal.

{
\subsection{Impact of imperfect cancellation}
\begin{figure}[tt]
\centering
\includegraphics[width = 8cm]{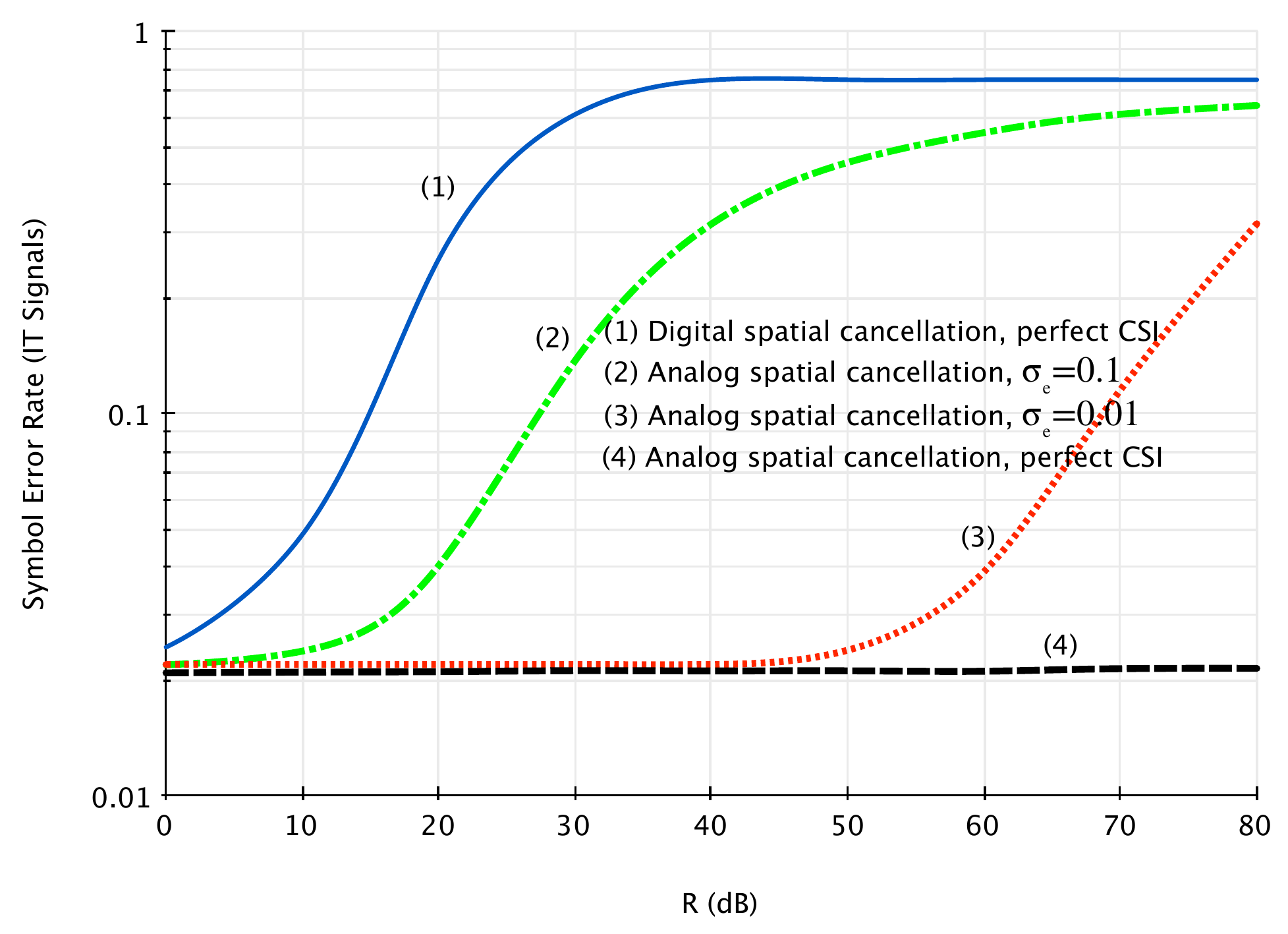}
\caption{Impact of imperfect cancellation on the SER performance of IT signal decoding.}
\label{Fig:imperfect cancellation}
\end{figure}

The impact of imperfect analog spatial cancellation on the SER performance of IT signal is investigated in Fig. \ref{Fig:imperfect cancellation}.  The imperfect factor is captured by a random perturbation $\Delta$ adding to the  of the phase shift elements of the cancellation matrix, i.e., $\tilde\Theta = \Theta + \Delta$, where $\tilde\Theta$ denotes the estimated phase shift parameter used in the cancellation matrix, while $\Theta$ is the actual phase shift as defined in (\ref{sys:1}).  The random perturbation can be incurred by practical impairments such as imperfect channel state information (CSI), phase noise and finite resolution of the implemented phase shifters, and is modelled as a additive Gaussian noise, following the Gaussian distribution ${\cal N}(0,\sigma_e^2)$ with zero mean and variance of $\sigma_e^2$. Two different levels of corruption are considered in this experiment, i.e., $\sigma_e = 0.01$ reflecting the mild CSI estimation error and $\sigma_e = 0.1$ representing the moderate CSI estimation error. In addition, the curves associated with perfect CSI aided digital and analog spatial cancellations are also plotted as benchmarks.
It can be observed that the performance of analog spatial cancellation is sensitive to the accuracy of the CSI estimation, especially in the large $R$ regime, e.g., $R = 80\;{\sf dB}$, where a mild error may still cause a relatively strong residual SWIPT interference degrading the SER performance of the IT signals. Nevertheless, it is also noted that although the performance of the analog spatial cancellation compromises from the imperfect CSI estimation, a decent performance gain can still be achieved even when moderate estimation error is occured, compared with the perfect CSI aided digital spatial cancellation, showing the superiority of the proposed analog cancellation technique in tackling the near-far problem.  
}


\subsection{Impact of the number of SWIPT BSs}
\begin{figure*}[tt]
  \centering
  \subfigure[$K=1$]{\label{fig:4a}\includegraphics[width=0.4\textwidth]{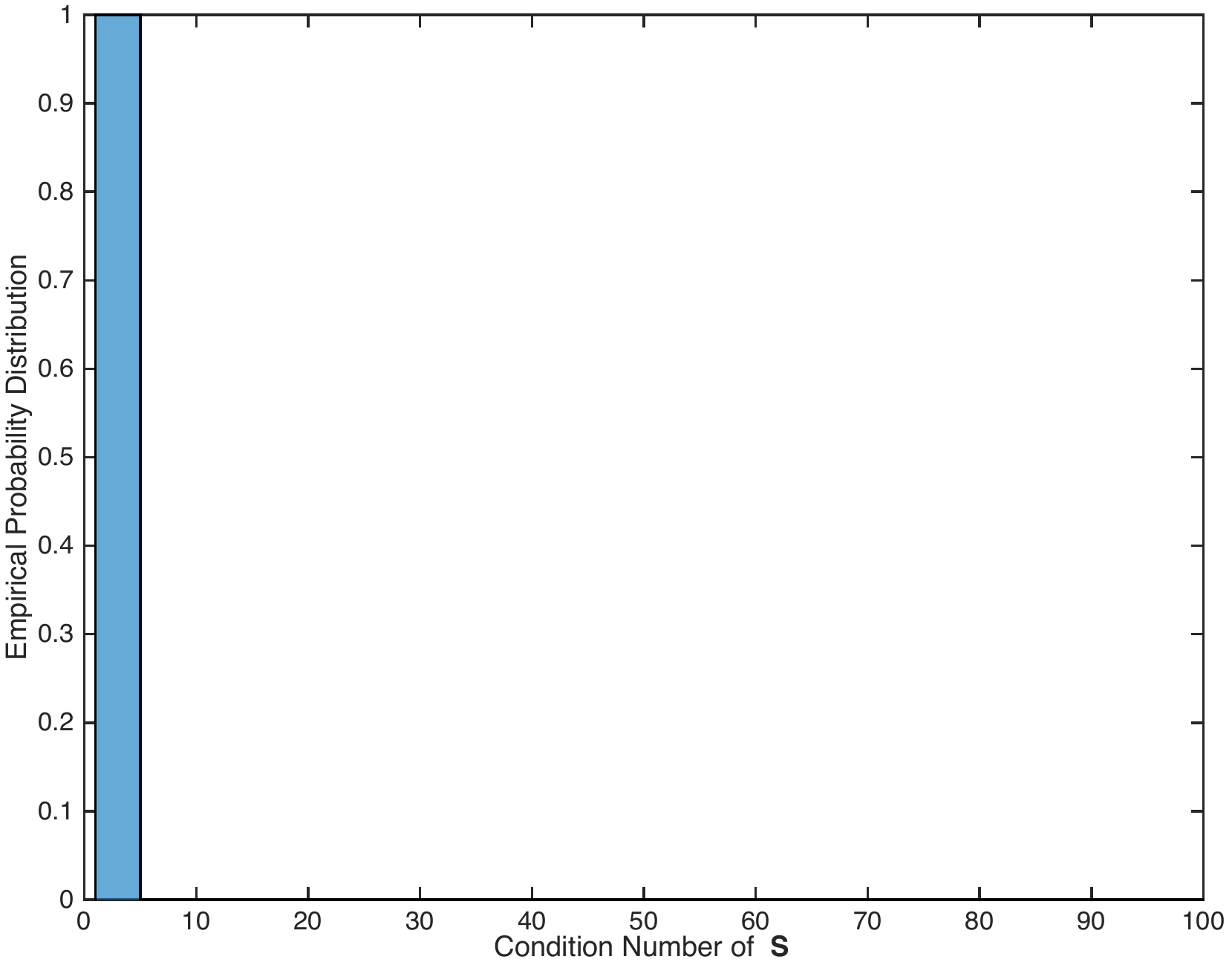}}
  \hspace{0.2in}
  \subfigure[$K=2$]{\label{fig:4b}\includegraphics[width=0.4\textwidth]{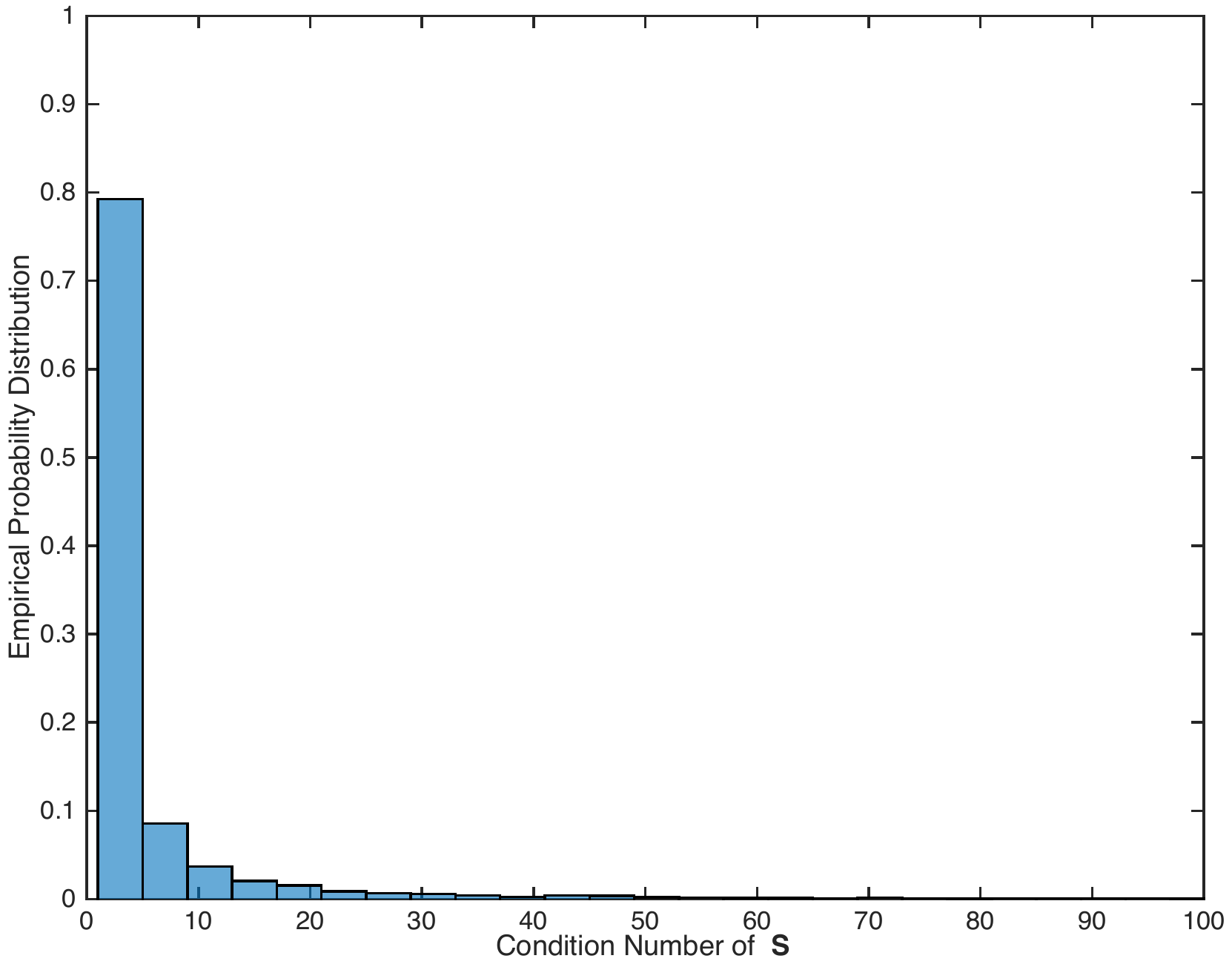}}
  \hspace{0.2in}
  \subfigure[$K=3$]{\label{fig:4c}\includegraphics[width=0.4\textwidth]{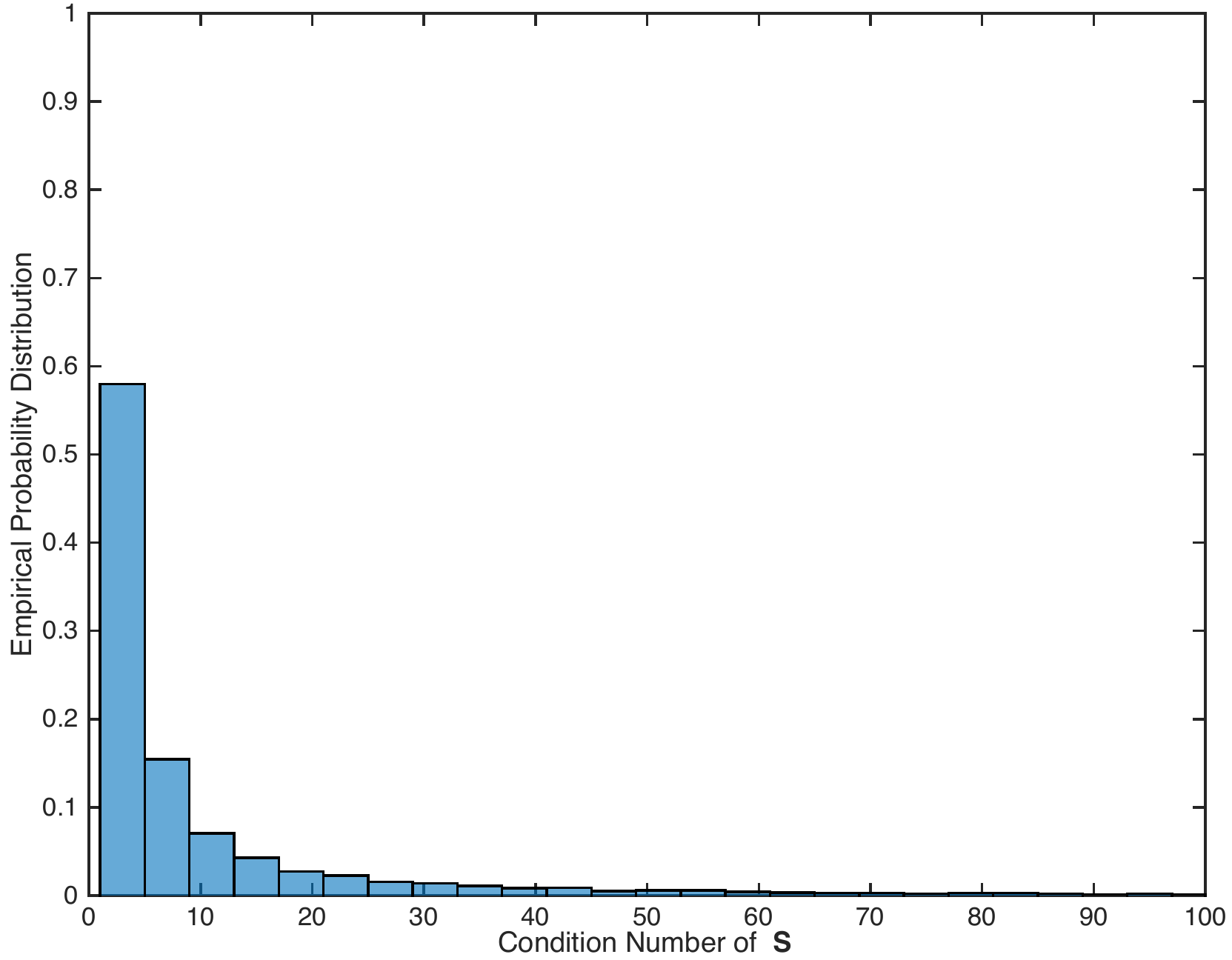}}
  \hspace{0.2in}
  \subfigure[$K=4$]{\label{fig:4d}\includegraphics[width=0.4\textwidth]{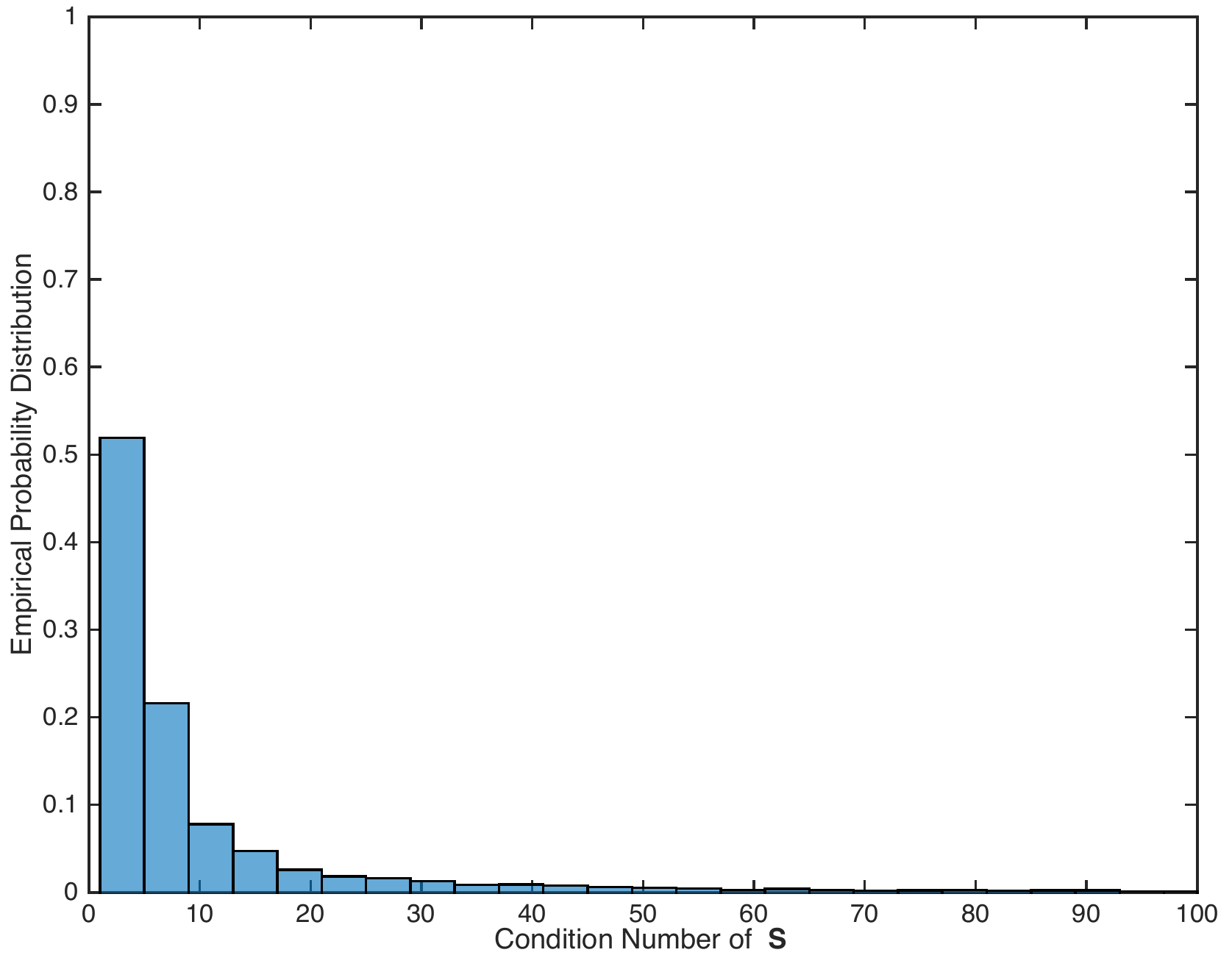}}
  \caption{The impact of $K$ of the condition number of cancellation matrix $\bf S$ for the case of $N_r = 16$.}
  \label{Fig:impact of K}
\end{figure*}

The impact of the number of SWIPT BSs on the condition number of $\bf S$ as revealed in Proposition \ref{prop:7} is verified by the simulation results in Fig. \ref{Fig:impact of K}. In this simulation, the number of receive antenna is set to be $N_r = 16$, and the optimal exhaust search is conducted to find out the desired cancellation matrix with the minimum condition number for the case of $K=2,3,4$. For all cases, the empirical probability distribution of the condition number of $\bf S$ is obtained using $10^6$ realizations of the phase shift parameters \{$\Theta_i$\} distributed uniformly over $[0,2\pi]$. It is observed that, the minimum condition number of one can always be achieved by the proposed Fourier/Hadamard based approach for the case of $K=1$ regardless of the realization of $\Theta$, as pointed out by Remark \ref{remark:4}. As $K$ increases from $1$ to $4$, the constructed cancellation matrix tends to have a larger condition number, which aligns with the discussion presented in Remark \ref{remark:prop:7}. Nevertheless, it is also noted that, even for the $K=4$ case, the distribution of the condition number of $\bf S$ still focuses on the relatively low-value region with a rapid decay rate, showing the robustness of the proposed Kronecker based construction approach in dealing with the cases of multiple SWIPT BSs.

\section{Conclusion}
To address the PT-IT near-far problem in the considered heterogeneous WPC network with an macro BS performing IT and densely deployed small-cell BSs performing SWIPT to the users, a novel technique called analog spatial cancellation has been proposed in this paper, which solves the problem by decoupling the received mixed strong SWIPT and the weak IT signals at the user in the analog domain. Practical designs for implementing analog spatial cancellation are presented, which consist of simple components such as adders and phase shifters. In particular, for the single-SWIPT-BS setup, two simple but optimal schemes based on Fourier or Hadamard transform are proposed, while for the multiple-SWIPT-BS setup, the design problem is more challenging, and a systematic Kronecker based construction framework is proposed to tackle the problem. It is shown that maximum multiplexing gain can still be achieved in the presence of the strong SWIPT signals when the proposed analog spatial cancellation is employed. 

The proposed analog spatial cancellation technology is of a significant practical interest, as it allows the coexistence of the SWIPT small-cell BS and the traditional macro BS by effectively handling the PT-IT near-far problem, unleashing the power transfer potential of a small-cell BS. Moreover, it motivates several promising research direction such as 1) protocol design for realizing cooperation communications between the SWIPT small-cell and BS to achieve a higher transmission rate; 2) distributed energy beamforming using multiple SWIPT small-cells located around the energy constraint mobile device to enhance the energy transmission efficiency; 3) safety-aware energy transmission by exploiting the said distributed energy beamforming technique to form an ultra-sharp beam targeted at the mobile device with a weak enough beam side lope towards the user guaranteeing the safety.

\appendices

\section{Proof of Proposition \ref{lemma:5}}\label{appendix:lemma:5}
The orthogonality of the cancellation vectors generated by the FBC construction method can be proven by exploiting the Kronecker product structure of the generated vectors as shown in the following.

Let ${\bf s}_a$ and ${\bf s}_b$ denote two arbitrary cancellation vectors generated by the FBC construction method. According to the Kronecker based construction framework, ${\bf s}_a$ and ${\bf s}_b$ can be expressed  as Kronecker products of component phase-shift vectors as follows.
\begin{equation}
{\bf s}_a = {\bf s}_{a(1)}\otimes{\bf s}_{a(2)}\otimes \cdots \otimes{\bf s}_{a(K)},
\end{equation} 
\begin{equation}
{\bf s}_b = {\bf s}_{b(1)}\otimes{\bf s}_{b(2)}\otimes \cdots \otimes{\bf s}_{b(K)},
\end{equation}
where ${\bf s}_{a(i)}$ and ${\bf s}_{b(i)}$ are the component phase-shift vectors of ${\bf s}_a$ and ${\bf s}_b$, respectively. 

Note that, in the Kronecker based construction framework, each component phase-shift vector is constructed as a product of a Fourier row vector and a phase compensation matrix, which gives ${\bf s}_{a(i)}^T = {\bf f}_{i,l}^T{\bf R}_i$ and ${\bf s}_{b(i)}^T = {\bf f}_{i,k}^T{\bf R}_i$ with ${\bf f}_{i,l}^T$,  ${\bf f}_{i,k}^T$ and ${\bf R}_i$ having the same definition as in (\ref{mp:2}).
Thus, invoking the mixed-product property of Kronecker product, we have
\begin{equation}
{\bf s}_a^H{\bf s}_b = {\bf s}_{a(1)}^H{\bf s}_{b(1)}\otimes{\bf s}_{a(2)}^H{\bf s}_{b(2)}\otimes \cdots \otimes {\bf s}_{a(K)}^H{\bf s}_{b(K)}.
\end{equation}
It is noted that ${\bf s}_a^H{\bf s}_b = 0$ if one of the ${\bf s}_{a(i)}^H{\bf s}_{b(i)} = 0$, $i = 1,2,\cdots,K$, which is always true since ${\bf s}_{a(i)}^H{\bf s}_{b(i)} = {\bf f}_{i,l}^H{\bf R}_i^H{\bf R}_i{\bf f}_{i,k} = {\bf f}_{i,l}^H{\bf f}_{i,k} = 0$ holds at least for a single $i$ given that any two ${\bf s}_a$ and ${\bf s}_b$ are two different vectors generated by the FBC construction procedure (at least for a single $i$, ${\bf f}_{i,l}$ and ${\bf f}_{i,k}$ come from different rows of a truncated Fourier matrix).

To this end, it is proven that any two cancellation vectors generated by the FBC construction procedure are orthogonal.

\section{Proof of Lemma \ref{lemma:2}}\label{appendix:lemma:2}
Note that the original inequality in (\ref{SP:2}) can be alternatively expressed as follows,
\begin{multline}\label{A:7}
\frac{1}{n!}\frac{P_1P_2\cdots P_n}{(P_1-1)(P_2-1)\cdots (P_n-1)} - \\ \frac{1}{(n-1)!(P_1-1)(P_2-1)\cdots (P_n-1)} \leq 1.
\end{multline}

For notation convenience, let's define $a_n = \frac{1}{n!}\frac{P_1P_2\cdots P_n}{(P_1-1)(P_2-1)\cdots (P_n-1)}$ and
$b_n = \frac{1}{(n-1)!(P_1-1)(P_2-1)\cdots (P_n-1)}$, hence, the proof of (\ref{A:7}) is equivalent to proving that, given any $n \in \mathbb{Z}^+$, the following inequality holds,
\begin{align}\label{A:8}
a_n - b_n \leq 1.
\end{align}

Since $P_i \geq 2$, we have $\frac{P_i}{P_i-1}\leq 2$, $i = 1,2,\cdots,n$, which gives
\begin{align}\label{A:9}
a_n \leq \frac{2^n}{n!}.
\end{align}

Thus, it is easy to note that $\forall n \geq 4$, we have $a_n < 1$. Also note that $a_n,b_n > 0$, therefore, it is proven that, for the case of $n \geq 4$, $a_n - b_n \leq 1$ holds. Now, to finish the whole proof, let's consider the rest cases of $n = 1,2,3$, respectively.

For $n = 1$, we have $a_1 = \frac{P_1}{P_1-1}$ and $b_1 = \frac{1}{P_1-1}$. Apparently, $a_1 - b_1 = 1$ satisfying (\ref{A:8}).

For $n = 2$, (\ref{SP:2}) reduces to $P_1P_2-2 \leq 2(P_1-1)(P_2-1)$ which can be alternatively given by
\begin{align}\label{A:10}
\frac{1}{2}(P_1-2)(P_2-2) \geq 0.
\end{align}
Note that $P_1,P_2 \geq 2$, hence, (\ref{A:10}) is certainly true, and the case for $n=2$ is proven.

Last, let's look into the case of $n=3$. A close observation reveals that $a_3 > 1$ only holds when $P_1 = P_2 = P_3 =2$. In other words, as long as not all $P_i \equiv 2$, $i=1,2,3$, we have $a_3 \leq 1$, thereby, $a_3 - b_3 < 1$ holds. As for the subcase that $P_1 = P_2 = P_3 =2$, it is easy to calculate that $a_3 = \frac{4}{3}$ and $b_3 = \frac{1}{2}$, which gives $a_3 - b_3 = \frac{5}{6} < 1$. As such, the proof for the case of $n=3$ is completed.

To this end, Lemma \ref{lemma:2} is proven.

\section{Proof of Lemma \ref{lemma:4}}\label{appendix:lemma:4}
In the following, we will prove the inequality using a special induction method called \emph{forward-backward induction} which is proposed by Augustin-Louis Cauchy for proving the well-known Cauchy--Schwarz inequality \cite{Bradley:Cauchy}.

For notation convenience, let's define the original inequality given in (\ref{NPB:2}) as $Q_m$ and it is equivalent to the following one.
\begin{align}\label{A:11}
\sqrt[m]{(P_1 - 1)(P_2 - 1)\cdots (P_m - 1)} \leq \sqrt[m]{P_1P_2\cdots P_m} - 1.
\end{align}

For $k=1$, obviously the equality holds.

For $k=2$, (\ref{A:11}) reduces to $(P_1-1)(P_2-1) \leq (\sqrt{P_1P_2} - 1)^2$ which can be alternatively expressed as
\begin{align}\label{A:12}
P_1P_2-P_1-P_2+1 \leq P_1P_2 - 2\sqrt{P_1P_2} + 1.
\end{align}
Note that $P_1+P_2 \geq 2\sqrt{P_1P_2}$, hence (\ref{A:12}) holds and $Q_2$ is proven.

Forward induction: If $Q_m$ holds, we can prove $Q_{2m}$ holds as follows,
\begin{align}
&\sqrt[2m]{(P_1 - 1)\cdots (P_m - 1)(P_{m+1} - 1)\cdots (P_{2m} - 1)} \notag\\
= &\sqrt{\sqrt[m]{(P_1 -1)\cdots (P_m - 1)} \sqrt[m]{(P_{m+1} - 1)\cdots (P_{2m} - 1)}} \notag\\
\leq &\sqrt{(\sqrt[m]{P_1 \cdots P_m} -1) (\sqrt[m]{P_{m+1}\cdots P_{2m}}-1) } \label{A:13}\\
\leq &\sqrt{\sqrt[m]{P_1\cdots P_m}\sqrt[m]{P_{m+1}\cdots P_{2m}} } - 1  \label{A:14}\\
= &\sqrt[2m]{P_1\cdots P_mP_{m+1}\cdots P_{2m}} -1 \notag ,
\end{align}
where (\ref{A:13}) holds because of $Q_m$ holds, and (\ref{A:14}) holds due to $Q_2$ holds.

Backward induction: If $Q_m$ holds, we can prove $Q_{m-1}$ holds as follows,

Let $A_m = \sqrt[m]{(P_1 - 1)\cdots (P_m -1)}$ and $B_m = \sqrt[m]{P_1\cdots P_m} - 1$.

Since $Q_m$ holds, we have
\begin{multline}\label{A:15}
\sqrt[m]{(P_1 - 1)\cdots (P_{m-1} - 1) A_{m-1}} \leq \\ \sqrt[m]{P_1\cdots P_{m-1}(A_{m-1} + 1)} - 1.
\end{multline}

Note that $A_{m-1}^{m-1} = (P_1 - 1)\cdots (P_{m-1} - 1)$ and $(B_{m-1} + 1)^{m-1} = P_1\cdots P_{m-1}$, hence (\ref{A:15}) can be rewritten as
\begin{align}\label{A:16}
&A_{m-1} \leq \sqrt[m]{(B_{m-1} + 1)^{m-1}(A_{m-1} + 1)} - 1 \notag\\
\Leftrightarrow &(A_{m-1} + 1)^m \leq (B_{m-1} + 1)^{m-1}(A_{m-1} + 1) \notag\\
\Leftrightarrow &(A_{m-1} + 1)^{m-1} \leq (B_{m-1} + 1)^{m-1} \notag\\
\Leftrightarrow &A_{m-1} \leq B_{m-1}
\end{align}

To this end, invoking $Q_2$ and forward induction we can prove $Q_{2^n}$, $n \in \mathbb{Z}^+$ holds, then further utilizing the backward induction we can prove that $\forall m$, $Q_m$ holds, which completes the proof.


\bibliographystyle{ieeetr}
\bibliography{BibDesk_File}

\vspace{-30pt}
\begin{IEEEbiography}[{\includegraphics[width=1in,clip,keepaspectratio]{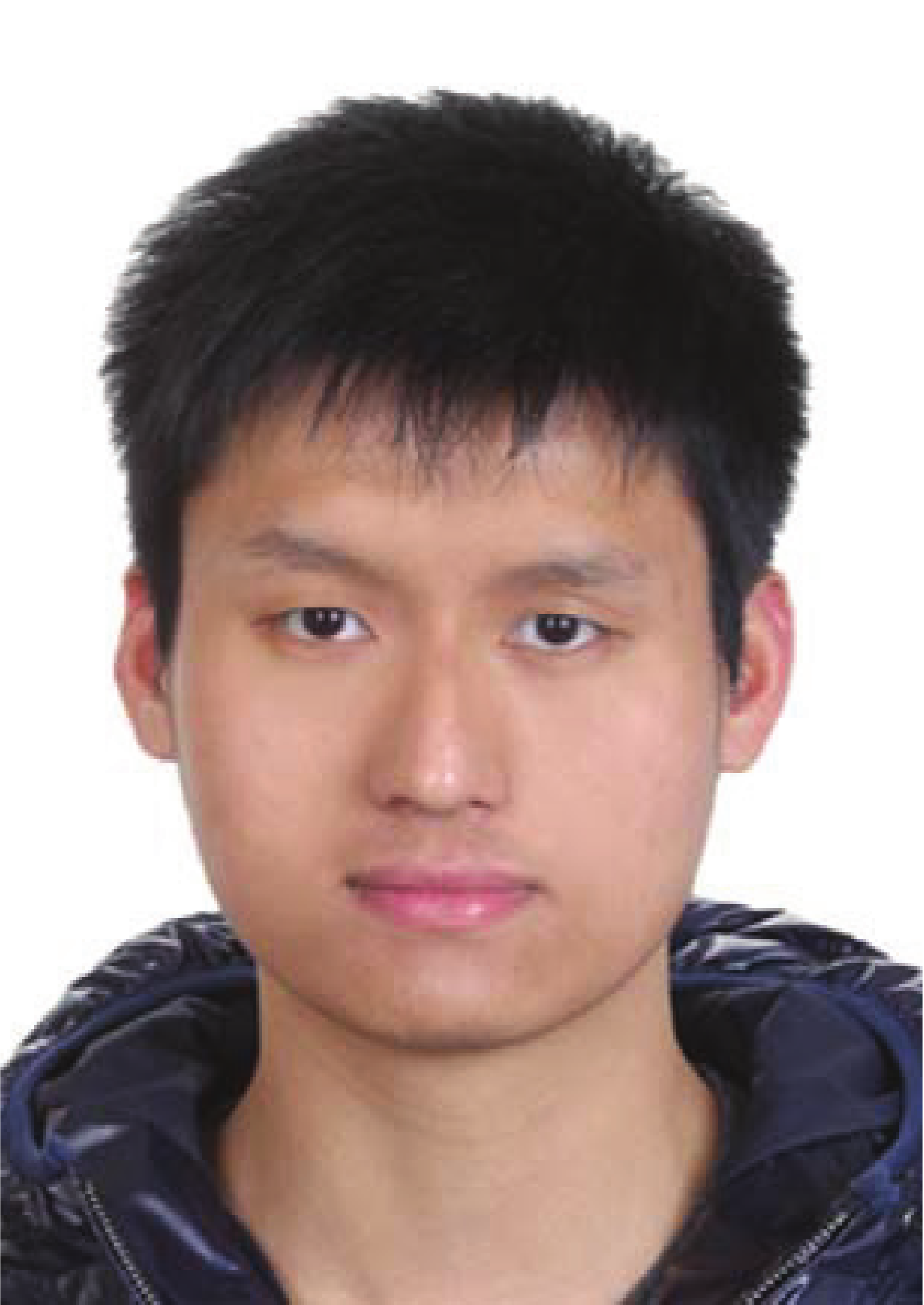}}]{Guangxu Zhu}
(S'14) received his B.S. and M.S. degree in Information and Communication Engineering from the Zhejiang University in 2012 and 2015, respectively. He is currently working towards his Ph.D. degree in the the department of EEE at the University of Hong Kong. His research interests include MIMO communications systems, cooperative communications and wirelessly powered communications. He is the recipient of a Best Paper Award from WCSP 2013.
\end{IEEEbiography}
\vspace{-30pt}
\begin{IEEEbiography}[{\includegraphics[width=1in,clip,keepaspectratio]{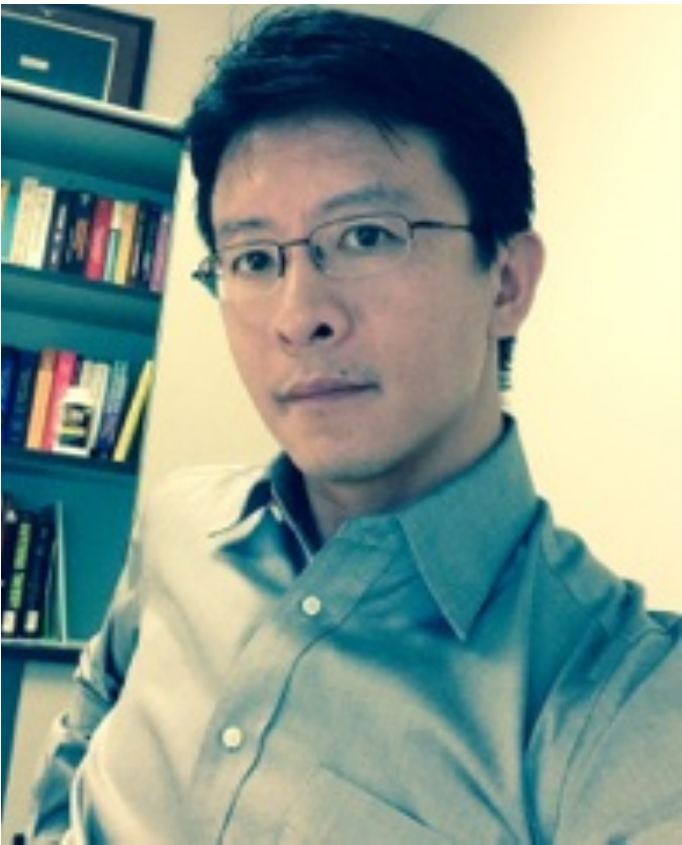}}]{Kaibin Huang}
(S'05, M'08, SM'13) received his Ph.D. degree from the University of Texas at Austin in electrical engineering. Since January 2014, he has been an assistant professor in the Department of EEE at the University of Hong Kong. He is an editor for IEEE JSAC Series on Green Communications and Networking, IEEE Transactions on Wireless Communications and also IEEE Wireless Communications Letters. He received a Best Paper Award from IEEE GLOBECOM 2006 and an IEEE Communications Society Asia Pacific Outstanding Paper Award in 2015. His research interests focus on the analysis and design of wireless networks using stochastic geometry and multi-antenna techniques.
\end{IEEEbiography}

\end{document}